\documentclass[10pt,journal,compsoc]{IEEEtran}

\usepackage{graphicx}
\graphicspath{{./}{Figures/}{res/}}
\usepackage{multirow}
\usepackage{subfigure}
\usepackage{amsfonts}
\usepackage{amsmath}
\usepackage{amssymb}
\usepackage{pifont}
\usepackage{colortbl}
\usepackage{cite}
\usepackage{bbding}
\usepackage{xcolor}
\usepackage{diagbox}
\usepackage{algorithmicx}
\usepackage{algorithm}
\usepackage{algpseudocode}

\usepackage{booktabs}
\usepackage{arydshln}
\usepackage{ulem}
\usepackage{bm}

\makeatletter
\newcommand{\thickhline}{%
    \noalign {\ifnum 0=`}\fi \hrule height 1pt
    \futurelet \reserved@a \@xhline
}
\newcolumntype{"}{@{\vrule width 1pt}}
\makeatother

\usepackage[colorlinks, citecolor=blue]{hyperref}
\newcommand{\tabincell}[2]{\begin{tabular}{@{}#1@{}}#2\end{tabular}}

\newcommand{\response}[1]{{\textcolor{black}{#1}}}

\makeatletter
 \newcommand*{\@rowstyle}{}
\newcommand*{\rowstyle}[1]{
 \gdef\@rowstyle{#1}%
 \@rowstyle\ignorespaces%
}
\newcolumntype{=}{
>{\gdef\@rowstyle{}}%
}
\newcolumntype{+}{
>{\@rowstyle}%
}
\newcolumntype{C}[1]{>{\centering\arraybackslash}p{#1}}
\makeatother

\ifCLASSINFOpdf
\else
\fi

\begin{document}

\title{Uformer-ICS: A U-Shaped Transformer for Image Compressive Sensing Service}

\author{Kuiyuan~Zhang,
        Zhongyun~Hua,
        Yuanman~Li,
        Yushu~Zhang,\\
        and Yicong~Zhou,~\IEEEmembership{Senior Member,~IEEE}
\IEEEcompsocitemizethanks{\IEEEcompsocthanksitem Kuiyuan~Zhang and Zhongyun~Hua are with the School of Computer Science and Technology, Harbin Institute of Technology, Shenzhen, Shenzhen 518055, China (e-mail: zkyhitsz@gmail.com; huazyum@gmail.com).
\IEEEcompsocthanksitem Yuanman~Li is with the College of Electronics and Information Engineering, Shenzhen University, Shenzhen 518060, China (e-mail: yuanmanli@szu.edu.cn).
\IEEEcompsocthanksitem Yushu~Zhang is with the College of Computer Science and Technology, Nanjing University of Aeronautics and Astronautics, Nanjing, Jiangsu 210016, China (e-mail: yushu@nuaa.edu.cn).
\IEEEcompsocthanksitem Yicong~Zhou is with the Department of Computer and Information Science, University of Macau, Macau 999078, China (e-mail: yicongzhou@um.edu.mo).
\IEEEcompsocthanksitem Corresponding author: Zhongyun Hua.}}


\IEEEtitleabstractindextext{%

\begin{abstract}
   %

   Many service computing applications require real-time dataset collection from multiple devices, necessitating efficient sampling techniques to reduce bandwidth and storage pressure. Compressive sensing (CS) has found wide-ranging applications in image acquisition and reconstruction. Recently, numerous deep-learning methods have been introduced for CS tasks.
   However, the accurate reconstruction of images from measurements remains a significant challenge, especially at low sampling rates.  
   In this paper, we propose Uformer-ICS as a novel U-shaped transformer for image CS tasks by introducing inner characteristics of CS into transformer architecture. 
   To utilize the uneven sparsity distribution of image blocks, we design an adaptive sampling architecture that 
   allocates measurement resources based on the estimated block sparsity, allowing the compressed results to retain maximum information from the original image.
    Additionally, we introduce a multi-channel projection (MCP) module inspired by traditional CS optimization methods. 
    By integrating the MCP module into the transformer blocks, we construct projection-based transformer blocks, and then form a symmetrical reconstruction model using these blocks and residual convolutional blocks.
    Therefore, our reconstruction model can simultaneously utilize the local features and long-range dependencies of image, and the prior projection knowledge of CS theory. 
    Experimental results demonstrate its significantly better reconstruction performance than state-of-the-art deep learning-based CS methods.
 \end{abstract}

}


\maketitle

\section{Introduction}
\IEEEPARstart{S}{ince} the advent of the information age, people have increasingly relied on the Internet to transmit, process, and share multimedia data, thereby gradually spurring the development of service computing. The cloud platform provides a seamless way to process and save data through the Internet. However, some service computing applications require real-time data collection from multiple devices~\cite{KaurRealTime2023,LiService2023}, necessitating sufficient network bandwidth and storage space. As a result, it's vital to improve the efficiency of data sampling and maintain the data content to reduce network bandwidth and storage space requirements for these applications. 
Compressive sensing (CS) is an effective signal acquisition technique that can reconstruct a signal using its compressed measurements, which are significantly less than the measurements required by the Nyquist sampling theorem~\cite{CS_Candes_2006,CS_Donoho_2006}. The sampling process for a signal $\mathbf{x} \in \mathbb{R}^{n \times 1}$ can be expressed as $\mathbf{y} = \mathbf{\Phi}\mathbf{x}$, where $\mathbf{\Phi}\in \mathbb{R}^{m\times n}$ is a measurement matrix wherein $n >> m$ and $\mathbf{y}\in \mathbb{R}^{m \times 1}$ comprises the compressed measurements. Because the CS technique can achieve a high compression ratio while maintaining a high reconstruction quality, it can be potentially used in numerous signal acquisition and compression applications~\cite{SCI-2021,YaoDimensionReduction2015}, especially for tasks involving image signals with high data redundancy~\cite{CS-Encryption-2019}.
The reconstruction process of CS involves finding solutions for an underdetermined linear system. 
To achieve high efficiency and accurate reconstruction results, most traditional CS reconstruction algorithms use nonlinear iterations to reconstruct the original signal~\cite{ISTA-2009,TV-2009,BCS-2007,BCS-FOCUSS_2017}, which is extremely time-consuming. Therefore, these traditional CS methods may not be suitable for real-time applications. Additionally, these traditional CS methods cannot achieve a high reconstruction quality because they lack adaptability to sample signals with varying characteristics.

Some deep learning-based image CS methods have applied convolutional neural networks (CNNs) in the CS task and can achieve both high reconstruction quality and efficiency~\cite{CSNet+_2020,SCSNet-2019,AMP-Net-2021,DPA-Net,OPINE-Net}. These deep learning-based image CS methods use CNN to replace the time-consuming iterative reconstruction process in traditional CS algorithms. They adaptively learn the measurement matrix and reconstruction network using a massive amount of training data. To maintain theoretical interpretability, some of them unfold the traditional CS algorithms into their CS networks using CNNs~\cite{AMP-Net-2021,OPINE-Net,you2021ISTA}. However, Owing to the spatial invariance and local inductive biases, CNNs cannot effectively capture long-range dependencies. This results in a limitation to these deep learning-based image CS methods. 

Recently, some CS methods~\cite{song2023optimization_OCTUF, shenTransCSTransformerBasedHybrid2022_TransCS, TCS_Net_2023,CSformer-2023} have employed the transformer architecture into CS tasks. The transformer has been extremely successful in various language processing~\cite{vaswani2017attention,XiaoLoader2023,YuBrain2023} and computer vision~\cite{liang2021swinir,ViT} tasks owing to the robust ability of its self-attention mechanism for capturing long-range dependencies. Inspired by the successful application of transformer, these CS methods ~\cite{song2023optimization_OCTUF, shenTransCSTransformerBasedHybrid2022_TransCS, TCS_Net_2023,CSformer-2023} utilize the transformer architecture for reconstruction and obtain better performance than previous CNN-based CS methods. While some transformer-based methods~\cite{shenTransCSTransformerBasedHybrid2022_TransCS,song2023optimization_OCTUF} also 
unfold the traditional CS algorithms, their image-level or single-channel level projection cannot make full use of multi-channel feature information in the middle stage. Furthermore, these methods~\cite{song2023optimization_OCTUF, shenTransCSTransformerBasedHybrid2022_TransCS, TCS_Net_2023,CSformer-2023} samples all the image blocks at the same sampling ratio without considering the unbalanced block sparsity. These limited correlations between inner CS characteristics and the transformer architecture hinder the further improvement of reconstruction performance.


In this work, we propose a specialized U-shaped transformer architecture for image CS called the Uformer-ICS, which introduces adaptive sampling and projection into the transformer architecture. 
A natural image usually has uneven sparsity distribution. 
To utilize the uneven sparsity distribution of image blocks, 
we design an adaptive sampling architecture that estimates sparsity from image measurements and allocates measurement resources based on the estimated block sparsity. The adaptive sampling model initially samples each image block at a small sampling ratio and estimates the block sparsity according to the measurements. Then it allocates fewer sampling resources to the image block with higher sparsity and vice versa. This approach can make the compressed measurements contain the maximum possible information of the sampled image under a fixed sampling ratio. The image signal is then adaptively sampled block-by-block using a single learnable measurement matrix. 
Additionally, we introduce a multi-channel projection (MCP) module inspired by traditional CS optimization methods~\cite{ISTA-2009,AMP-2009}. The original projection operation is applied block-wise on the image domain. To make it suitable for transformer architecture and make full use of multi-channel information, the MCP module adapts projection on the multi-channel feature domain. By integrating the MCP module into the transformer blocks, we construct projection-based transformer blocks, and then build a symmetrical reconstruction model using these blocks and residual convolutional blocks. Therefore, our reconstruction model can simultaneously utilize the local features, and long-range dependencies of image and the prior projection knowledge of CS theory. Our Uformer-ICS is an end-to-end framework that simultaneously learns the sampling and reconstruction processes. Experimental results demonstrate its superior reconstruction performance compared to existing state-of-the-art deep
learning-based CS methods.

\begin{figure}[!t]
  \centering
    \includegraphics[width=0.95\linewidth]{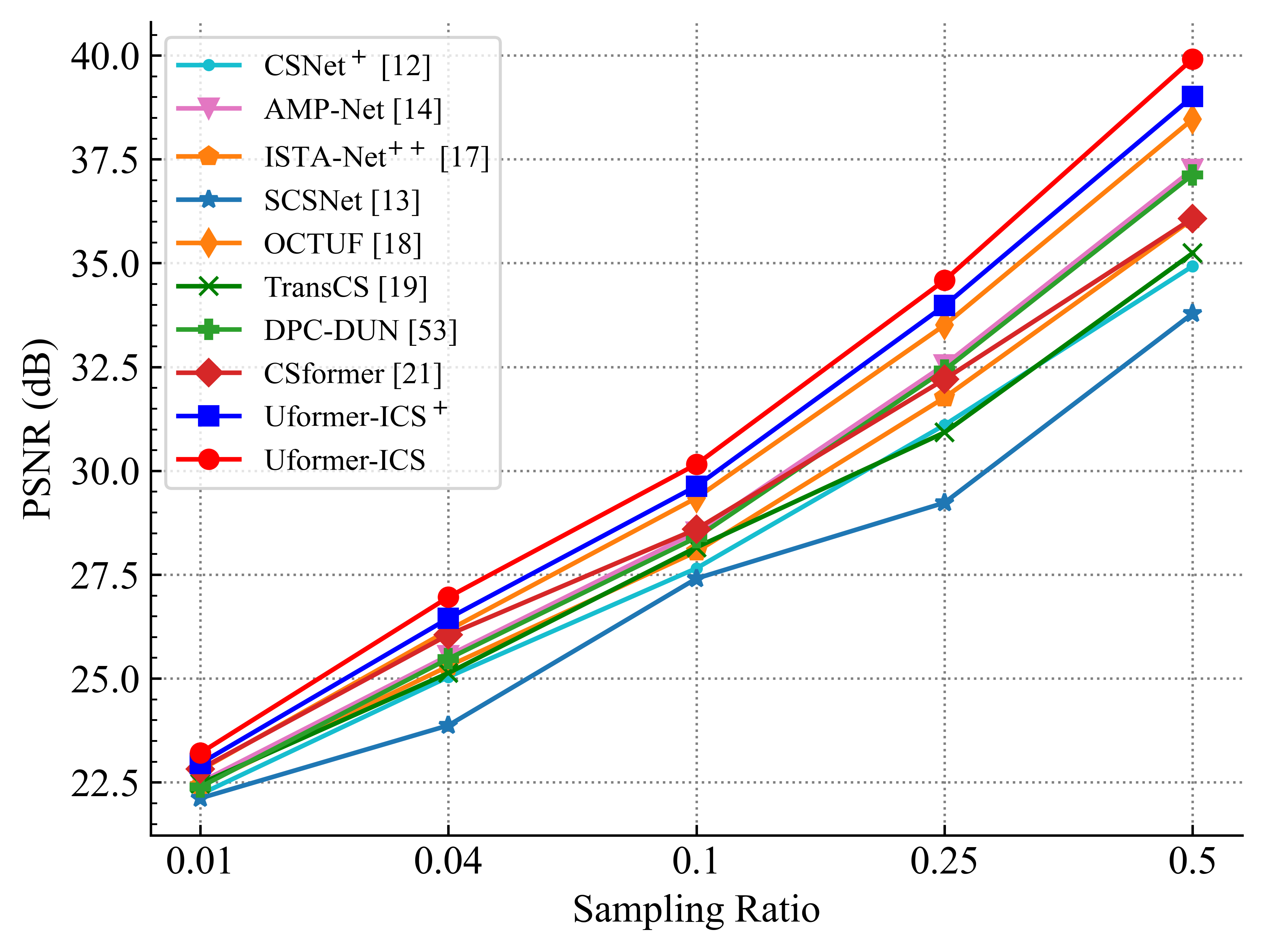}
      \vspace{-10pt}
    \caption{Average reconstruction performances of the proposed and existing state-of-the-art deep learning-based CS methods. The peak signal-to-noise ratio (PSNR) scores shown are averaged over all images in the five test datasets: Set5, Set11, Set14, BSD100, and Urban100. It is evident that the proposed method achieves significantly better PSNR scores than state-of-the-art deep learning-based CS methods.}
    \label{Fig.avg_PSNR}
\end{figure}

The main contributions of this study are as follows:
\begin{itemize}
  \item We propose a new transformer-based network for image CS, called the Uformer-ICS, which effectively introduces two CS characteristics, adaptive sampling and projection, into the transformer architecture. 

 \item We design an adaptive sampling architecture to allocate sampling resources to image blocks based on block sparsity. The block sparsity is estimated from the initial measurements of the sampled image, and we evaluate three sparsity estimation methods in the adaptive sampling architecture.
 
  \item  We design a MCP module to adapt the projection operation into the multi-channel feature domain, and then develop a symmetrical reconstruction model using a projection-based transformer and residual convolutional blocks, wherein each projection-based transformer block is constructed by integrating the MCP module into the original transformer blocks.
    
  \item We conduct extensive experiments to evaluate the proposed Uformer-ICS. The comparison results demonstrate that it outperforms existing state-of-the-art deep learning-based CS methods (as shown in Fig.~\ref{Fig.avg_PSNR}), and the ablation studies validate the effectiveness of the proposed adaptive sampling strategy and the MCP module.
  
\end{itemize}

The rest of this paper is organized as follows. Section~\ref{Sec.related_work} reviews the traditional CS methods, deep learning-based CS methods, and vision transformers. Section~\ref{Sec.method} presents the network structure of the proposed Uformer-ICS. Section~\ref{Sec.experiments} evaluates the performance of the proposed method and compares it with state-of-the-art methods.  Finally, conclusions are drawn in Section~\ref{Sec.conclusion}.

\section{Related Work}
\label{Sec.related_work}
In this section, we present some knowledge of the traditional CS methods, and review existing deep learning-based CS methods and vision transformers. 

\vspace{-6pt}
\subsection{Traditional CS}
\subsubsection{Block-Based CS}
The size of the measurement matrix in CS is calculated by multiplying the lengths of the input signal and output measurements. When directly sampling the 1D vector form of a 2D image, the required measurement matrix is considerably large, which results in significantly high memory and computational requirements for the sampling operation. To address this issue, traditional CS algorithms usually sample an image block-by-block~\cite{BCS-2007}.  Suppose that each image block $\mathbf{x}_i$ is of size $B\times B$ and the measurement matrix is $\mathbf{\Phi}_B \in \mathbb{R}^{n_B \times B^2}$, where $n_B$ is the number of sampled measurements. The sampling process is expressed as
\begin{equation}
  \mathbf{y}_i = \mathbf{\Phi}_B \cdot \mathfrak{T}(\mathbf{x}_i),
  \label{Eq_BCS}
\end{equation}
where $\mathfrak{T}(\mathbf{x}_i)$ denotes the 1D vector form of the image block $\mathbf{x}_i$ and $\mathbf{y}_i\in \mathbb{R}^{n_B \times 1}$ are the measurements of $\mathbf{x}_i$. The sampling ratio ($sr$) is defined as $sr = n_B/B^2$. Generally, an image block with higher sparsity requires fewer measurements for reconstruction~\cite{yingyuSaliencyBasedCompressiveSampling2010}. Therefore, some traditional CS methods improve the reconstruction quality by allocating sampling resources to different blocks~\cite{fowlerMultiscaleBlockCompressed2011,chenSelfadaptiveSamplingRate2014}.

\subsubsection{Signal Reconstruction}
To effectively reconstruct the original signal, some traditional CS methods have developed iterative reconstruction schemes to solve the following optimization problem:
\begin{equation}
\min _{\hat{\mathbf{x}}_i} \frac{1}{2}\|\mathbf{\Phi}_B \mathfrak{T}(\hat{\mathbf{x}}_i)-\mathbf{y}_i\|_{2}^{2}+ \mathfrak{R}(\hat{\mathbf{x}}_i),
\label{Eq.CSReconstruction}
\end{equation}
where $\mathfrak{R}(\hat{\mathbf{x}}_i)$ denotes the hand-crafted prior term regarding the structure of the original signal~\cite{ISTA-2009,TV-2009,AMP-2009,BCS-FOCUSS_2017}. Each iteration operation includes the following operations:
\begin{itemize}
    \item \textbf{Projection onto the convex set.} Given the reconstruction result $\hat{\mathbf{x}}_i^{(t)}$ of the $t$-th iteration step, this operation uses the gradient descent method to minimize $\|\mathbf{\Phi}_B \mathfrak{T}(\hat{\mathbf{x}}_i)-\mathbf{y}_i\|_{2}^{2}$  in Eq.~\eqref{Eq.CSReconstruction}. Therefore, it can find a vector $\hat{\mathbf{x}}_i^{(t+1)}$ closer to the hyperplane $\mathbf{H}=$ $\left\{\hat{\mathbf{x}}_i: \mathbf{\Phi}_B \mathfrak{T}(\hat{\mathbf{x}}_i)=\mathbf{y}_i\right\}$ than $\hat{\mathbf{x}}_i^{(t)}$ without a constraint, e.g., Sparsity constraint. Specifically, the projection operation is calculated as follows: 
    \begin{equation}
    \mathfrak{T}(\hat{\mathbf{x}}_i^{(t+1)}) = \mathfrak{T}(\hat{\mathbf{x}}_i^{(t)}) + \mathbf{\Phi}_B^{T}(\mathbf{y}_i - \mathbf{\Phi}_B \mathfrak{T}(\hat{\mathbf{x}}_i^{(t)})) / (1 + \alpha ), 
    \label{Eq_projection}
    \end{equation}
    where  $\mathbf{\Phi}_B^{T}(\mathbf{y}_i - \mathbf{\Phi}_B \mathfrak{T}(\hat{\mathbf{x}}_i^{(t)}))$ denotes the gradient of $\|\mathbf{\Phi}_B \mathfrak{T}(\hat{\mathbf{x}}_i)-\mathbf{y}_i\|_{2}^{2}$ in Eq.~\eqref{Eq.CSReconstruction} and $\alpha$ is the updating step length.
    
    \item \textbf{Optimizing the prior term.} This operation aims to minimize the hand-crafted prior term $\mathfrak{R}(\hat{\mathbf{x}}_i)$ to constrain the current reconstruction result $\hat{\mathbf{x}}_i^{(t+1)}$. For example, the iterative shrinkage-thresholding algorithm (ISTA)~\cite{ISTA-2009} and approximate message passing algorithm (AMP)~\cite{AMP-2009} set the $l_1$ norm as the prior term to constrain the signal Sparsity, and use a shrinkage/thresholding non-linearity to process $\hat{\mathbf{x}}_i^{(t+1)}$. The denoising-based AMP (DAMP)~\cite{DAMP-2016} assumes that the original signal belongs to a certain image class $\mathcal{C}$, and uses a denoiser to project $\hat{\mathbf{x}}_i^{(t+1)}$ onto $\mathcal{C}$.
    
\end{itemize}
Despite having good interpretability, traditional CS algorithms suffer from poor reconstruction performance and low efficiency because they cannot adaptively learn the features of the original signal.

\begin{figure*}[!htbp]
  \centering
    \includegraphics[width=0.95\linewidth]{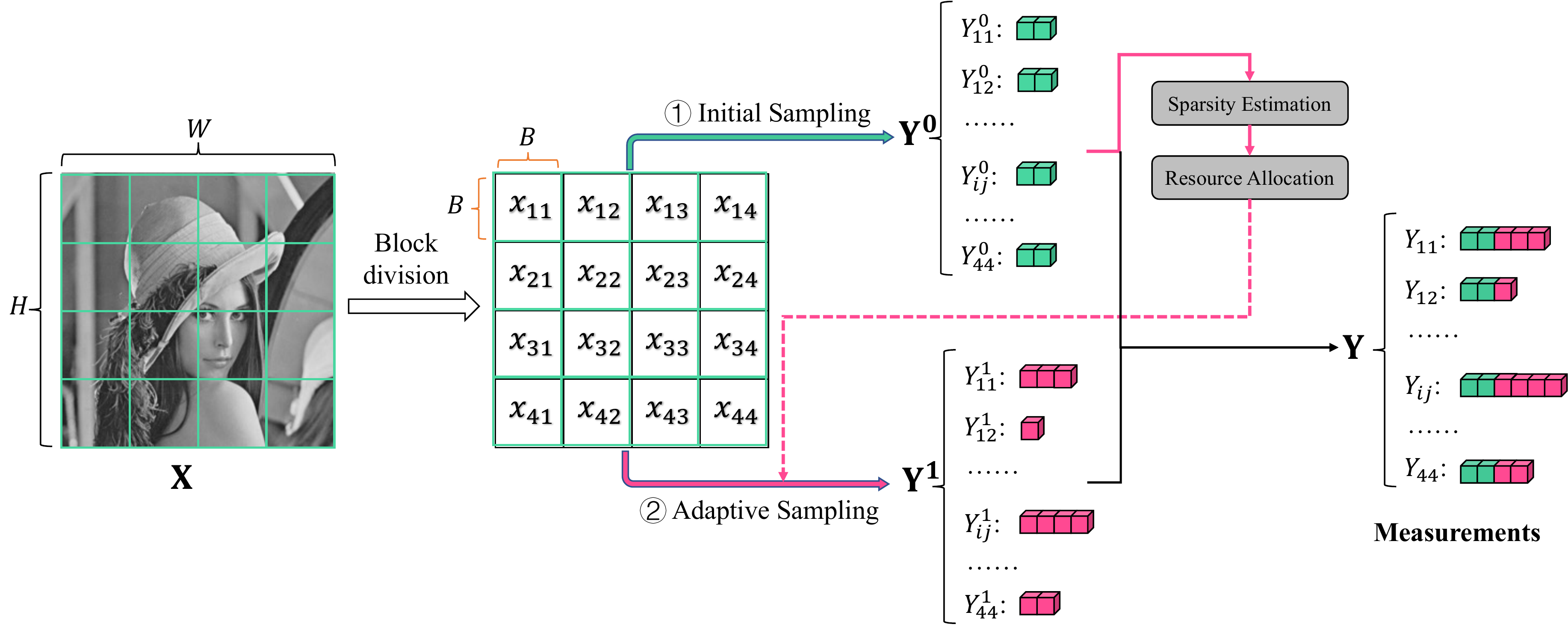}
      \vspace{-6pt}
    \caption{Adaptive sampling model of the proposed Uformer-ICS. First, the image $\mathbf{X}$ is initially sampled, and the initial measurements are utilized to estimate the block sparsity, which is used to adaptively allocate sampling resources for each block. Then, the image is further adaptively sampled block-by-block. The final measurements are obtained by concatenating the initial measurements and adaptive measurements.}
    \label{Fig.sampling}
\end{figure*}

\subsection{Vision Transformer}

In 2017, the authors of~\cite{vaswani2017attention} first proposed the transformer for natural language processing tasks. Compared with previous methods based on recurrent neural networks, the transformer has a more powerful ability to model long-range dependencies among tokens using the self-attention mechanism;thus, it can achieve significantly better accuracy and scalability. Specifically, the self-attention layer computes the key-query dot-product among all input tokens. Therefore, its computation complexity grows quadratically with an increase in the number of tokens~\cite{vaswani2017attention}. 

To efficiently handle high-resolution images, Vaswani $et~al.$~\cite{ViT} proposed the vision transformer (ViT) that divides an image into non-overlapping patches and employs the transformer to capture dependencies among image patches for image classification tasks. In some low-level vision tasks (e.g., image denoising and image deblocking), an image may be divided into numerous patches, making the computational complexity of ViT extremely high and unacceptable. Considering this, Liu $et~al.$~\cite{liuSwinTransformerHierarchical2021} designed a shifted window-based transformer,  which applies self-attention in each image window and uses shifted window partitioning to bring connections across windows. Moreover, Zamir $et~al.$~\cite{zamir2021restormer} proposed a transposed attention scheme that computes the attention map across feature channels. These efforts~\cite{liuSwinTransformerHierarchical2021,zamir2021restormer} have significantly reduced the computational complexity of the transformer for image processing tasks, achieving a linear relationship between the computational complexity and the image size.


\subsection{Deep Learning-Based CS}
In 2015, the authors of~\cite{SDA-2015} used a deep learning method to solve the signal reconstruction problem and proposed a stacked denoising auto-encoder to reconstruct image patches from their measurements. Inspired by the application of CNNs in image restoration tasks, Kulkarni $et~al.$~\cite{ReconNet_2016} proposed a deep CNN-based model to implement the non-iterative reconstruction process. These two deep learning-based CS methods directly reconstruct image patches from their measurements~\cite{SDA-2015,ReconNet_2016}. To further improve the reconstruction performance, Shi $et~al.$ proposed two end-to-end deep CNN-based models in~\cite{CSNet*_2017,CSNet+_2020} for image CS, which can simultaneously learn both the sampling and reconstruction processes.

The above CS models directly regard the reconstruction process as a deep learning task without considering the characteristics of CS theory. To improve the interpretability, some studies~\cite{ISTA-Net_2018,AMP-Net-2021,OPINE-Net} designed CS networks by unfolding traditional CS algorithms using CNNs. For example, inspired by the traditional ISTA method~\cite{ISTA-2009},  Zhang $et~al.$~\cite{ISTA-Net_2018} proposed a deep network called ISTA-Net by solving the proximal mapping associated with the sparsity-inducing regularizer using nonlinear transformation. The authors of~\cite{AMP-Net-2021} proposed a denoising-based deep CS network called AMP-Net, wherein the reconstruction network unfolds the iterative denoising process of the AMP algorithm~\cite{AMP-2009} and integrates deblocking modules to eliminate the blocking artifacts.

Recently, some works~\cite{song2023optimization_OCTUF,shenTransCSTransformerBasedHybrid2022_TransCS, TCS_Net_2023, CSformer-2023} have employed the transformer architecture on the CS tasks inspired by the successful application of the transformer on vision tasks. The method~\cite{TCS_Net_2023} modified the transformer architecture to fit the patch-to-pixel multi-stage pattern and built a transformer-based hierarchical framework for CS tasks called TCS-Net. Ye $et~al.$~\cite{CSformer-2023} proposed a CNN-Transformer hybrid framework to explore the representation capacity of local and global features. To introduce CS characteristics into the transformer architecture, the method~\cite{shenTransCSTransformerBasedHybrid2022_TransCS} builds an ISTA-based transformer backbone that iteratively works with projection operation, and the method~\cite{song2023optimization_OCTUF} fuses the projection operation into a cross attention block. However, they utilize the projection on the image-level or single-channel level, which cannot fully use multi-channel feature information in the middle stage. Moreover, the sparsity in different areas of a natural image usually varies, and these methods~\cite{song2023optimization_OCTUF,shenTransCSTransformerBasedHybrid2022_TransCS, TCS_Net_2023, CSformer-2023} sample all image blocks using the same sampling ratio without appropriately considering the uneven sparsity distribution. These insufficient correlations between CS characteristics and the transformer architecture cause a limit to getting further improvement in the reconstruction performance, especially at small sampling ratios.


\section{Uformer-ICS}
\label{Sec.method}
This section details the proposed Uformer-ICS. From an overall perspective, the Uformer-ICS comprises an adaptive sampling model and a reconstruction model. The adaptive sampling model adaptively samples the image block-by-block using a single learnable measurement matrix, and the reconstruction model is a symmetrical U-shaped architecture that reconstructs the original image from the measurements. 

To illustrate the sampling and reconstruction processes, we assume that the sampled image is a single-channel image that is denoted as $\mathbf{X}\in \mathbb{R}^{H\times W\times 1}$, where $H$ and $W$ are the height and width of image, respectively,  and the block size is  $B\times B \times 1$. The vector form of each image block $\mathbf{x}_{ij}$ is represented as $\mathfrak{T}(\mathbf{x}_{ij})\in \mathbb{R}^{B^2\times 1}$. Herein, $i=\{1, 2,\cdots,h\}$ and $j=\{1, 2,\cdots,w\}$, where $h=\frac{H}{B}$ and $w=\frac{W}{B}$.

\subsection{Adaptive Sampling Model}
The detailed illustration of the adaptive sampling model is shown in Fig.~\ref{Fig.sampling}. The adaptive sampling model first samples the initial measurements for each image block, then estimates the block sparsity and assign sampling resources, and finally adaptively samples the image.


\subsubsection{Initial sampling} 
It is impossible to access the image before sampling and reconstructing it in some real sampling scenarios. In works~\cite{BCS-Net_2019,ChenContent2022}, the sampling networks directly apply transforms on the image to calculate saliency before sampling it. 
Therefore, they are limited to tasks that have access to the image before sampling, such as image compression and re-sampling tasks, where the image is already stored on the disk or in memory.

Given the target sampling ratio $sr_t$ for the whole image, we initially sample each image block with a smaller sampling ratio $sr_{init}$ and then adaptively assign the rest sampling resources according to the initial measurements. Assume that the learnable measurement matrix is $\mathbf{\Phi}\in\mathbb{R}^{M\times B^2}$, where $M$ is the maximum number of measurements for each image block. Each image block $\mathbf{x}_{ij}$ is initially sampled as follows:
\begin{equation}
    \begin{aligned}
        n^0 &= \lfloor B^2 * sr_{init} \rfloor \\
        \mathbf{y}_{ij}^0 &= \mathbf{\Phi}[1:n^0, :]\cdot \mathfrak{T}(\mathbf{x}_{ij}),
    \end{aligned}
\end{equation}
where $n^0$ is the initial measurement number,  $\mathbf{y}_{ij}^0 \in\mathbb{R}^{n^0 \times 1}$ is the initial measurements of the image block $\mathbf{x}_{ij}$.

\subsubsection{Sparsity Estimation}
\label{sec.sparsity_estimation}
Because different image blocks have different sparsity, the compressed measurements cannot contain the maximum information of the original image if all image blocks are sampled at the same sampling ratio. Generally speaking, an image block with higher sparsity should be allocated fewer sampling resources because it contains less information.

Before using sparsity estimation methods, we first obtain the initial estimation $\mathbf{x}_{ij}^0$ of each image block:
\begin{equation}
    \begin{aligned}
        \mathbf{x}_{ij}^0 &= \mathbf{\Psi}[:, 1:n^0]\cdot \mathbf{y}_{ij}^0.
    \end{aligned}
\end{equation}
where $\mathbf{\Psi}\in\mathbb{R}^{B^2\times M}$ represents the learnable linear mapping matrix. By reshaping and concatenating all $\mathbf{x}_{ij}^0$, a low-quality estimation $\mathbf{X}^0$ is obtained for the input image $\mathbf{X}$.


To estimate the sparsity of the image blocks, we utilize three methods: saliency map (SM), standard deviation (STD), and block difference (DIFF).

\begin{algorithm}[!tbp]
  \caption{Sampling resources allocation.}
  \label{Alg.adptive_sampling}
  \begin{algorithmic}[1]
      \Require Sparsity information $\mathbf{V}$, target sampling ratio $sr_t$, image size $H\times W$, block size $B$, initial number $n^0$ of measurements 
      \Ensure measurement number $\mathbf{m}_{ij}$ 
      \State $h, w$ = $H/B, W/B$
      \State $total = \lfloor sr_t \times H \times W \rfloor$
      \State $rest = total - (n^0 \times h \times w)$
      \For{$i = 1 : h$}
        \For{$j = 1 : w$}
            \State $\mathbf{V}_{ij} = \mathbf{V}_{ij} /\operatorname{sum}(\mathbf{V})$
            \State $\mathbf{m}_{ij} =  n^0 + \lfloor rest \times \mathbf{V}_{ij} \rfloor$ 
        \EndFor
      \EndFor
    \end{algorithmic}
\end{algorithm}

\textbf{Saliency map}. The saliency map method employs discrete cosine transform (DCT) to calculate the saliency information~\cite{yingyuSaliencyBasedCompressiveSampling2010,BCS-Net_2019,ZhangAMS-Net2022}. Specifically, the sparsity information $\mathbf{V}_{ij}$ for each block $\mathbf{x}_{ij}$ is calculated as
\begin{equation}
    \begin{aligned}
    \mathbf{F} &=\operatorname{abs}(\mathrm{C_t}^{-1}(\operatorname{sign}(\mathrm{C_t}(\mathbf{X}^0)))), \\
    \mathbf{S} &=\mathbf{G} * \mathbf{F}^{2}, \\
    \mathbf{V}_{ij} &= \sum_{s \in \mathbf{S}_{\mathbf{x}_{ij}}} s / \sum_{s \in \mathbf{S}} s,
    \end{aligned}
\end{equation}
where $\mathrm{C_t}$ and $\mathrm{C_t}^{-1}$ denotes the 2D DCT and its reverse operation, $\mathbf{G}$ is a 2D Gaussian low-pass filter for smoothing, and $\mathbf{S}_{\mathbf{x}_{ij}^0}$ represents the corresponding region for each block $\mathbf{x}_{ij}^0$.

\textbf{Standard deviation.} The work~\cite{chen2014self_std} assumes that an image block $\mathbf{x}_{ij}$ with a larger standard deviation usually has a larger complexity and estimates the standard deviation through the measurements sampled by the Gaussian matrix. Inspired by it, we calculate the standard deviation of each low-quality estimation $\mathbf{x}_{ij}^0$ as the sparsity information $\mathbf{V}_{ij}$ for each block $\mathbf{x}_{ij}$.

\textbf{Block difference.} The adjacent regions in nature images usually have similar patterns, which benefit the image denoising in the reconstruction. Therefore, the image block having more similar adjacent blocks can be regarded as more sparse and allocated fewer sampling resources. For an image block $\mathbf{x}_{ij}^0$, we first calculate the absolute difference with four adjacent blocks $\{ \mathbf{x}_{i, j-1}^0$,$\mathbf{x}_{i-1, j}^0$, $\mathbf{x}_{i, j+1}^0$, $\mathbf{x}_{i+1, j+1}^0 \}$, and then average the difference into a scalar as the sparsity information $\mathbf{V}_{ij}$.


\subsubsection{Adaptive Sampling}
We assign the rest measurements using the sparsity information $\mathbf{V}_{ij}$ and apply adaptive sampling for each image block.
According to the sparsity estimation methods in Section~\ref{sec.sparsity_estimation}, the sparsity information $\mathbf{V}_{ij}$ is negatively related to the sparsity. In other words, a larger $\mathbf{V}_{ij}$ means that the corresponding image block $\mathbf{x}_{ij}$ is less sparse and should be allocated more sampling resources.


Algorithm~\ref{Alg.adptive_sampling} describes the process of allocating sampling resources. The total number of measurements under the target sampling ratio $sr_t$ are first calculated,  and the remaining sampling resources are linearly assigned for each image block $\mathbf{x}_{ij}$ based on the sparsity information $\mathbf{V}_{ij}$. Finally, the measurement number $\mathbf{m}_{ij}\in (1, B^2]$ is obtained for each image block $\mathbf{x}_{ij}$.



\begin{figure}[!tbp]
    \centering
    \begin{minipage}[b]{0.99\linewidth}
        \centering
        \begin{minipage}[b]{0.45\linewidth}
            \centerline{\includegraphics[width = 1\linewidth]{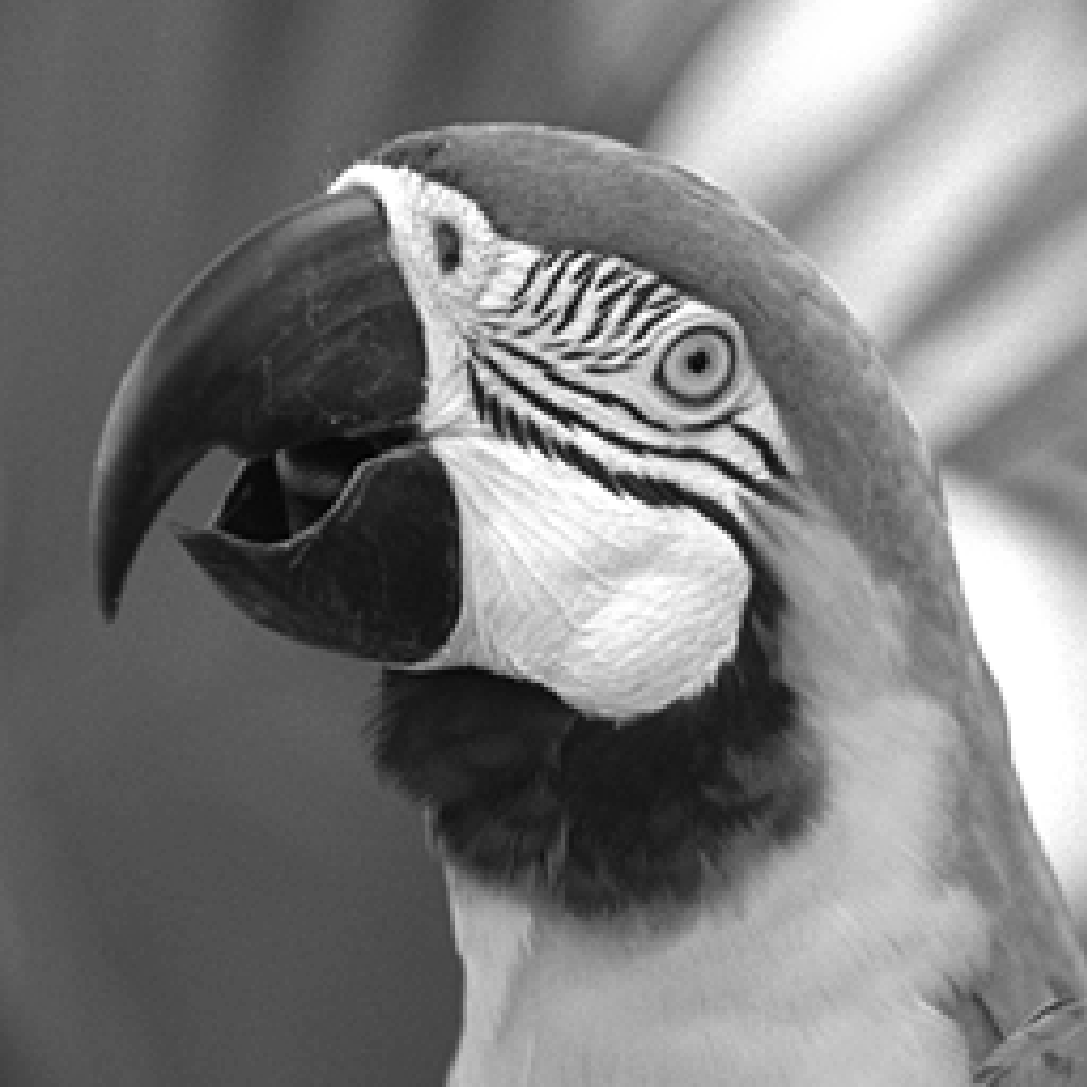}}
            \centerline{\footnotesize{(a) Original image}}
            \vspace{2pt}
            \centerline{\includegraphics[width = 1\linewidth]{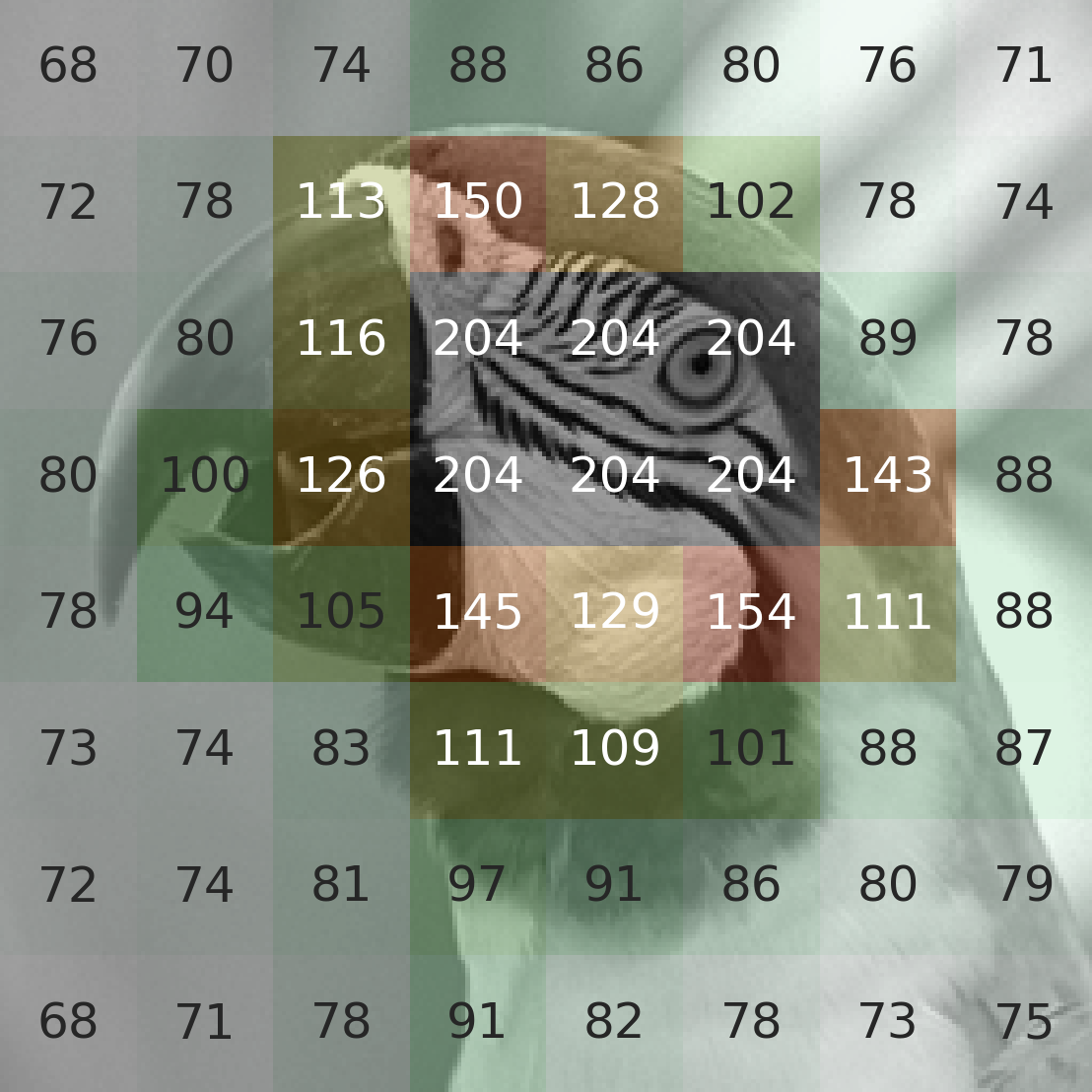}}
            \centerline{\footnotesize{(c) Saliency map}}
        \end{minipage}
        \begin{minipage}[b]{0.45\linewidth}
            \centerline{\includegraphics[width =1\linewidth]{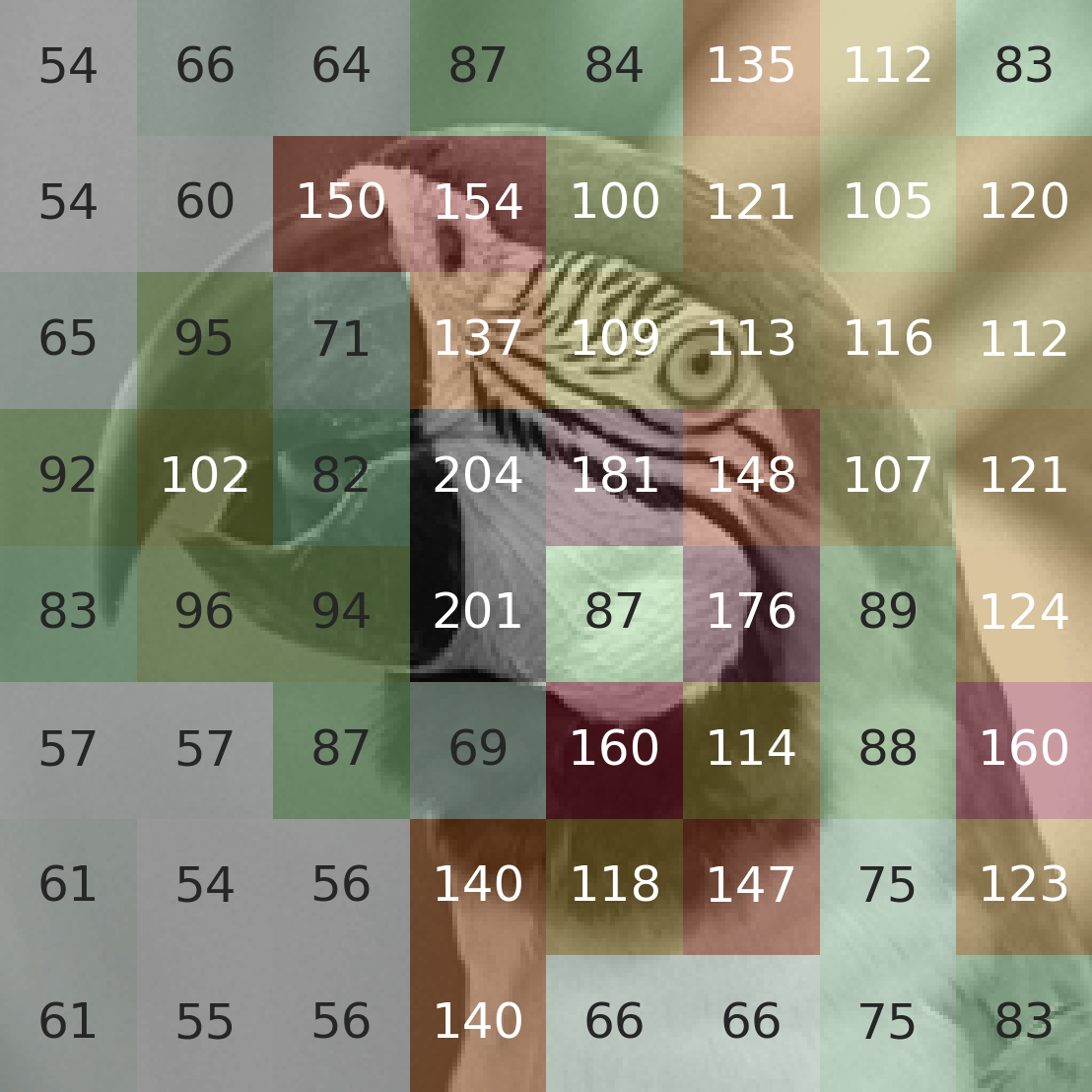}}
            \centerline{\footnotesize{(b) Standard deviation}}
            \vspace{2pt}
            \centerline{\includegraphics[width = 1\linewidth]{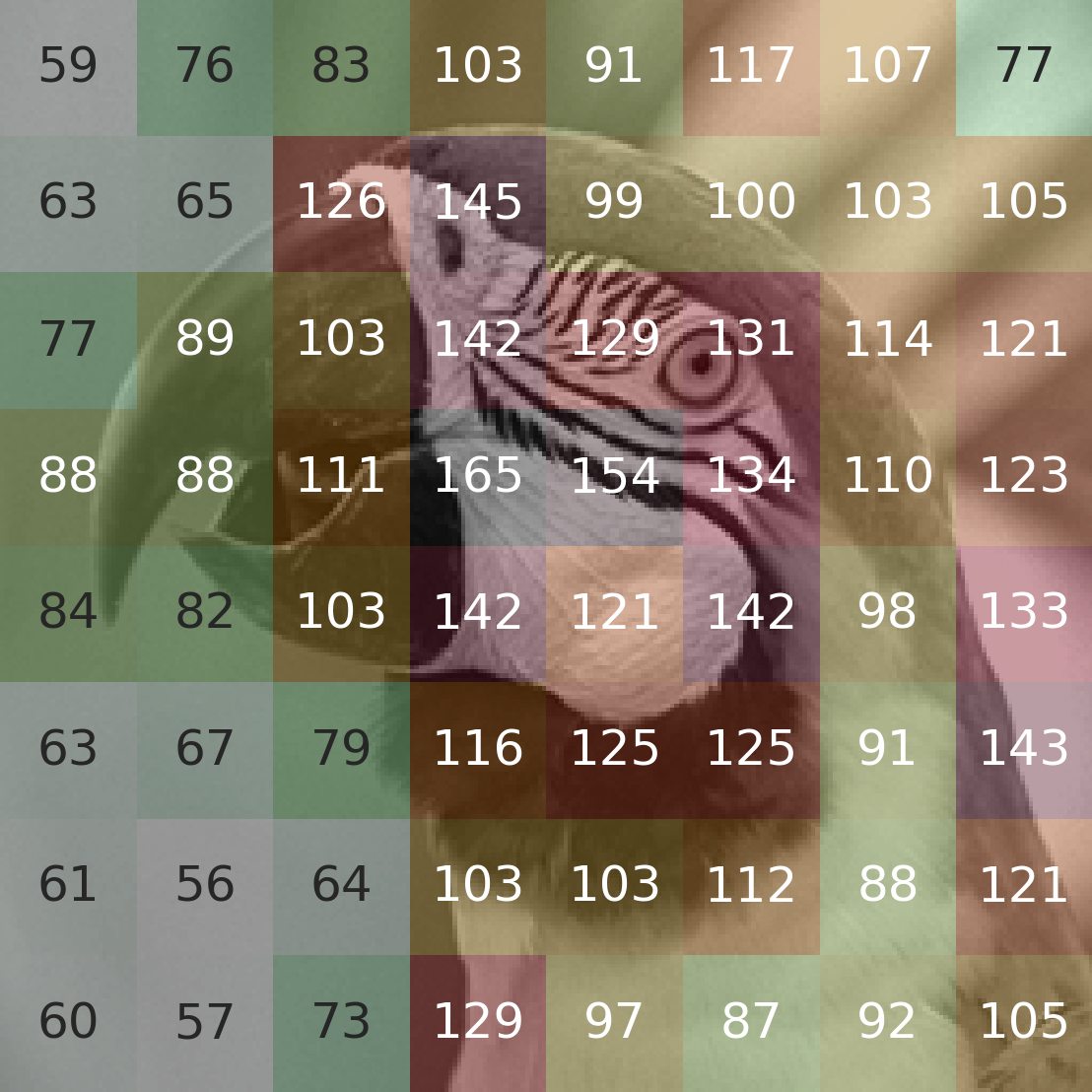}}
            \centerline{\footnotesize{(d) Block difference}}
        \end{minipage}
    \end{minipage}
    \caption{Measurement allocations using three sparsity estimation methods for the ``Parrots'' image at the sampling ratio of 0.1.}
    \label{Fig.measurement_map}
\end{figure}

To better illustrate the adaptive sampling mechanism, we provide one example of the measurement allocations using three sparsity estimation methods for the ``Parrots'' image shown in Fig.~\ref{Fig.measurement_map}. The block size is set to $32\times 32$,  and the number on the image block denotes the measurement number. It is evident that image blocks with more complex structures are considered to have less sparsity and are allocated more measurements. For example, the eye in the ``Parrots'' contains more details and more complex structures; thus, the corresponding image blocks in the eye area are allocated more sampling resources. The upper-left and lower-left corners are black backgrounds and therefore are allocated less sampling resources.

\begin{figure*}[!htbp]
  \centering
    \includegraphics[width=0.99\linewidth]{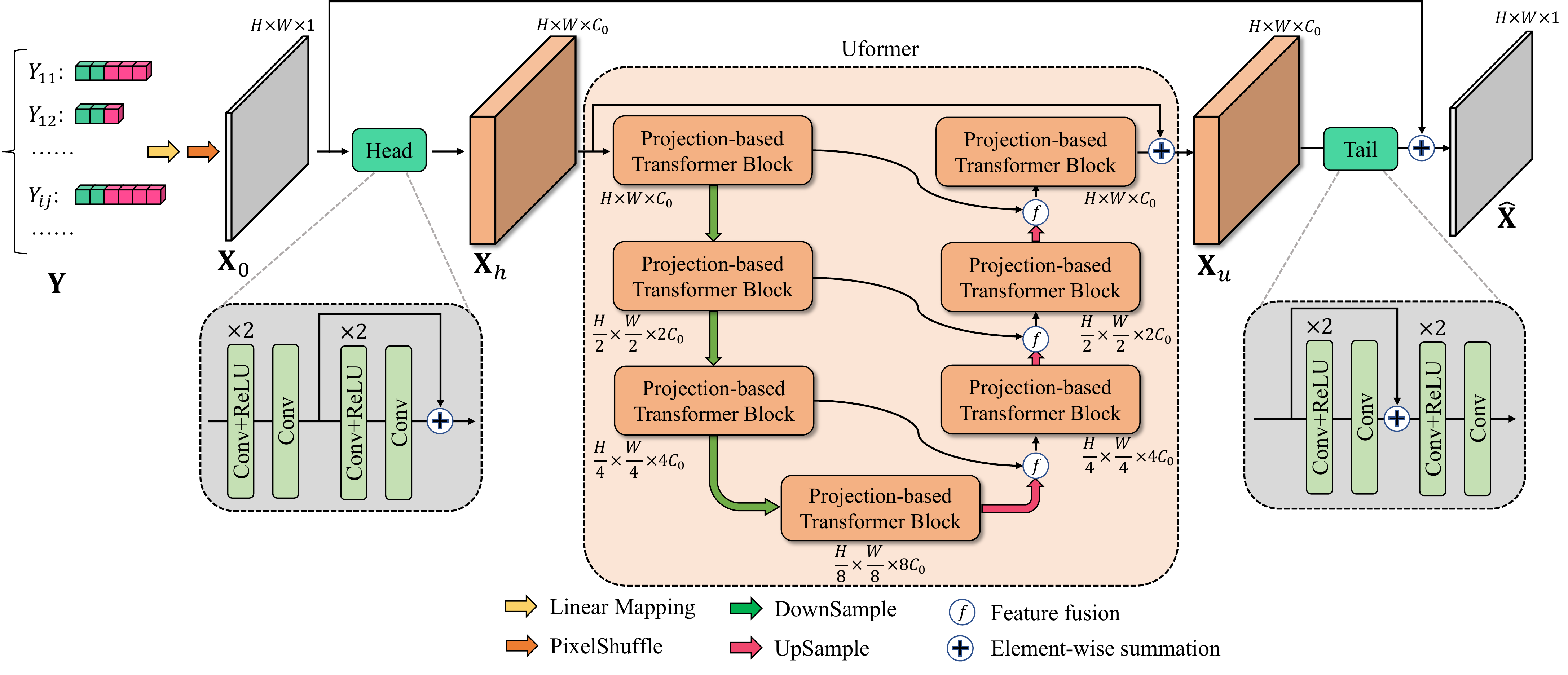}
      \vspace{-10pt}
    \caption{Overview of the reconstruction model of the proposed Uformer-ICS. Given the adaptive sampling result $\mathbf{Y}$, the reconstruction model first applies linear mapping on it and employs pixel shuffle operation to transform the combined mapping results into an image-like initialization $\mathbf{X}_0$. Then, the reconstruction model feeds the obtained initialization $\mathbf{X}_0$ into the \textit{Head} module for extracting shallow features $\mathbf{X}_h$. Taking $\mathbf{X}_h$ as input, the \textit{Uformer} module captures the long-range dependencies to enhance the feature representation and outputs $\mathbf{X}_u$. Finally, the \textit{Tail} module generates the final reconstruction result $\hat{\mathbf{X}}$ by adding the initialization $\mathbf{X}_0$ and aggregation features of $\mathbf{X}_u$.}
    \label{Fig.reconstruction}
\end{figure*}

Because different image blocks are allocated different measurement numbers, generating a measurement matrix for each possible measurement number results in a significant number of parameters and extremely high memory occupation. To address this problem, we use a scalable sampling strategy that requires only one learnable measurement matrix to sample all image blocks with different measurement numbers. Specifically, we reuse the learnable measurement matrix $\mathbf{\Phi}$ in the initial sampling. Each image block is sampled as follows:
\begin{equation}
    \begin{aligned}
    \mathbf{y}_{ij}^1 &= \mathbf{\Phi}[n^0+1:\mathbf{m}_{ij}, :]\cdot \mathfrak{T}(\mathbf{x}_{ij}),\\
    \mathbf{y}_{ij} &= \operatorname{concat}(\mathbf{y}_{ij}^0, \mathbf{y}_{ij}^1),
    \end{aligned}
\end{equation}
where $\mathbf{y}_{ij}\in\mathbb{R}^{\mathbf{m}_{ij} \times 1}$ is the measurements of the image block $\mathbf{x}_{ij}$. We denote the collection of all measurements $\mathbf{y}_{ij}$ as $\mathbf{Y}$.

\subsection{Reconstruction Model}

The overall structure of the reconstruction model is illustrated in Fig.~\ref{Fig.reconstruction}.
First, the initialization value $\mathbf{X_0}\in \mathbb{R}^{H\times W\times 1}$ of the input image $\mathbf{X}$ is generated by linearly mapping the measurements $\mathbf{Y}$ directly. Specifically, we reuse learnable linear mapping matrix $\mathbf{\Psi}$ in the sparsity estimation, and the image initialization is calculated as:
\begin{equation}
    \begin{aligned}
    \mathfrak{T}(\hat{\mathbf{x}}_{ij}) &= \mathbf{\Psi}[:, 1:\mathbf{m}_{ij}] \cdot \mathbf{y}_{ij} \\
    \mathbf{X_0} & = \Xi(\{\hat{\mathbf{x}}_{ij}\}),
    \end{aligned}
\end{equation}
where $\Xi$ denotes the pixel shuffle operation~\cite{PixelShuffle} that repositions elements from the channel dimension to the spatial dimension and we use it to transform the combined linear mapping results into an image-like feature  $\mathbf{X_0}$.

Next, the initialization $\mathbf{X}_0$ is processed using a head module $H_{head}$ to extend the channels and extract the local shadow features $\mathbf{X}_h \in \mathbb{R}^{H\times W\times C_0}$, where $C_0$ is the number of channels. The Uformer module then captures the long-range features from $\mathbf{X}_h$ to obtain a feature $\mathbf{X}_u$. Finally, a tail module $H_{tail}$ aggregates $\mathbf{X}_u$ to produce a residual reconstruction result, and the final reconstruction result $\hat{\mathbf{X}}$ is generated by adding the initialization $\mathbf{X}_0$ and the residual reconstruction result.


In the subsequent subsections, the head and tail modules are first illustrated and then the Uformer module is described in detail.

\begin{figure}[tbp]
  \centering
    \includegraphics[width=0.98\linewidth]{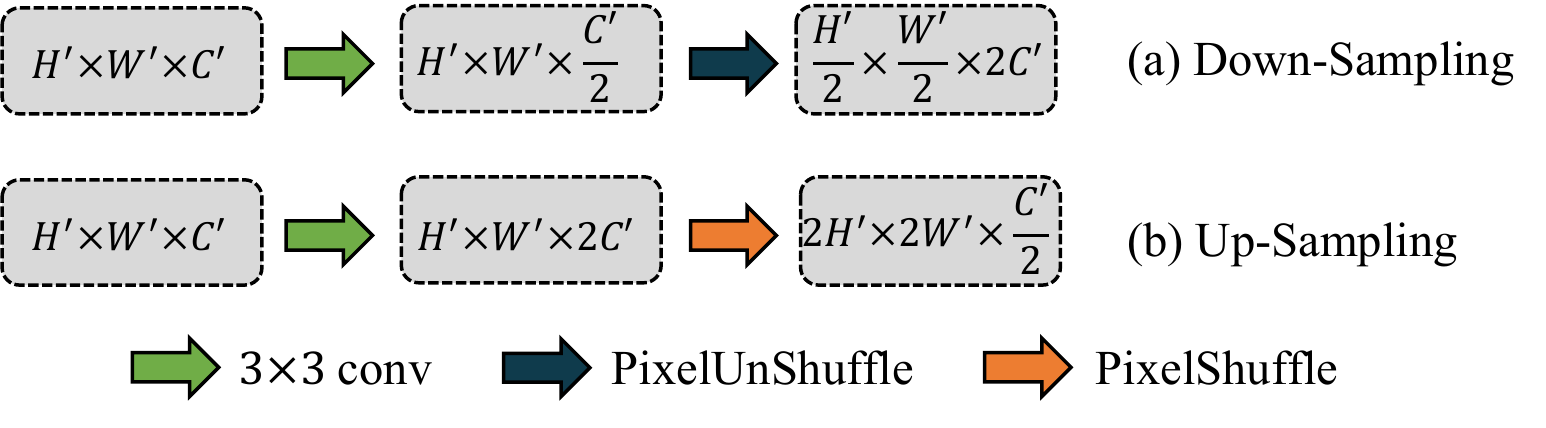}
      \vspace{-10pt}
    \caption{Illustrations of the feature down-sampling and up-sampling operations.}
    \label{Fig.up_down_proj}
\end{figure}

\subsection{Head \& Tail Modules}

As shown in Fig.~\ref{Fig.reconstruction}, the head module $H_{head}$ comprises several convolutional layers and a residual convolutional block. After obtaining the initialization result $\mathbf{X_0}$, the $H_{head}$ enlarges the channels further and extracts local shallow  features from $\mathbf{X_0}$ as follows:
\begin{equation}
    \mathbf{X}_{h} = H_{head}(\mathbf{X_0}),
\end{equation}
where $\mathbf{X}_{h}$ is of size $H\times W \times C_0$.

To ensure symmetry with the head module $H_{head}$, the tail module $H_{tail}$ stacks a residual convolutional block and several convolutional layers. Given output $\mathbf{X}_u$ of the Uformer module, the $H_{tail}$ aggregates the features of $\mathbf{X}_u$ to obtain the final reconstruction result $\hat{\mathbf{X}}$ as follows:
\begin{equation}
    \hat{\mathbf{X}} = H_{tail}(\mathbf{X}_u) + \mathbf{X_0},
\end{equation}
where $\hat{\mathbf{X}}$ is of size $H\times W \times 1$, and the residual learning~\cite{he2016deep} is applied to improve the convergence speed and reconstruction performance of the proposed model.

\subsection{Uformer}

As illustrated in Fig.~\ref{Fig.reconstruction}, the Uformer is a U-shaped four-level hierarchical encoder-decoder. To introduce the prior knowledge of CS into the transformer architecture, we propose a multi-channel projection (MCP) and integrate it into the stacked transformer blocks to develop the projection-based transformer block. In the following sections, we will illustrate the overall pipeline of the Uformer and present its components in detail.

\begin{figure*}[!tbp]
    \centering
    \centerline{\includegraphics[width = 0.85\linewidth]{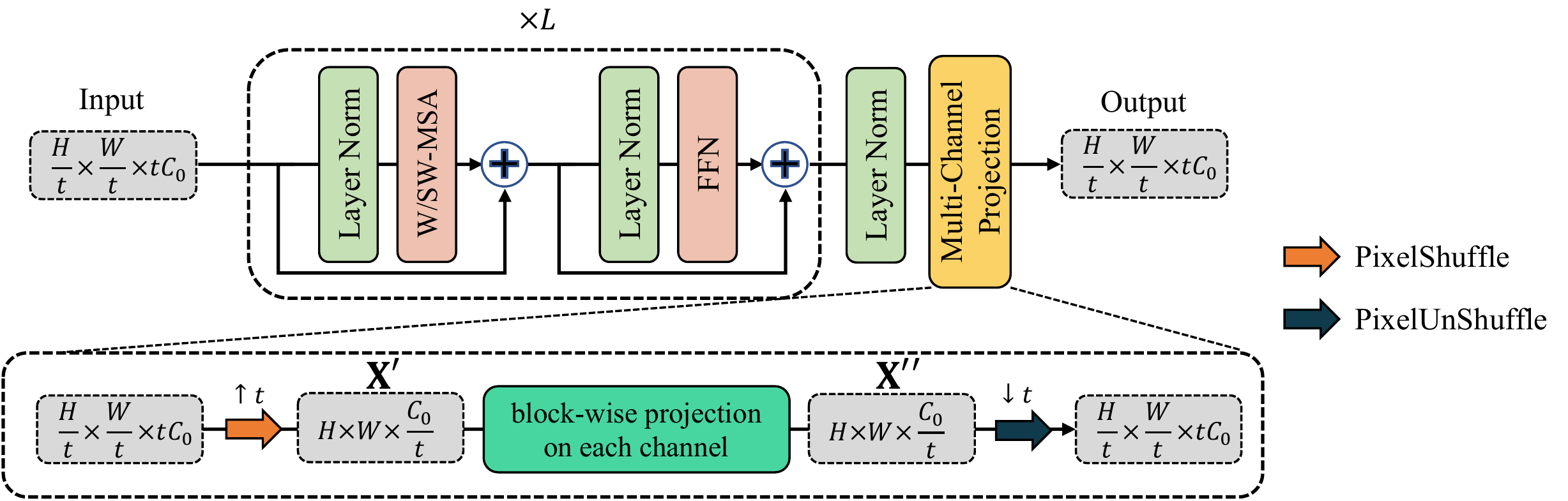}}
      \vspace{-10pt}
    \caption{Structure of the projection-based transformer block. First, $L$ Transformer blocks are stacked to learn long-range dependencies, and then the multi-channel projection module is used to update the multi-channel features block-wise.}
    \label{Fig.mc-projection}
\end{figure*}

\subsubsection{Pipeline}

Based on the common design of the U-shaped structure~\cite{ronneberger2015u}, the output $\mathbf{X}_h$ of the head module is first passed through four encoder levels. The resolutions of feature maps are gradually reduced from the top to the bottom levels using feature down-sampling operations shown in Fig.~\ref{Fig.up_down_proj} (a). We employ a $3\times 3$ convolutional layer and pixel unshuffle operation~\cite{PixelShuffle} to double the number of channels and reduce the resolution by $50\%$. Therefore, the $i$-th level of the encoder produces a feature map of size $\frac{H}{2^{i}}\times \frac{W}{2^{i}} \times 2^{i}C_0$, where $i\in [0, 1, 2, 3]$.

The decoder progressively recovers the high-resolution features by using the low-resolution feature map from the lowest encoder level as its input. Fig.~\ref{Fig.up_down_proj} (b) shows the feature up-sampling operation. We employ a $3\times 3$ convolutional layer and pixel shuffle operation~\cite{PixelShuffle} to reduce the channels by $50\%$ and double the resolution. Moreover, the features obtained on the decoder side are fused with the encoder features to enhance the feature representation.



\subsubsection{Window-Based Self-attention}

The computational complexity of the original self-attention layer in the transformer increases quadratically with image size. In this study, the window-based multi-head self-attention (W-MSA)~\cite{liuSwinTransformerHierarchical2021} is used to reduce the computational complexity because it computes self-attention in non-overlapping windows and its computational complexity increases linearly with image size. Given an input feature map $\mathbf{Z}\in \mathbb{R}^{H'\times W' \times d}$, the W-MSA divides it into non-overlapping windows of size $W\times W \times d$. Suppose that $\mathbf{P}\in \mathbb{R}^{W^2 \times d}$ denotes the $W^2$ pixels in a window, the  W-MSA applies the self-attention process to these pixels as follows:
\begin{equation}
\begin{aligned}
    \mathbf{Q, K, V} &= \mathbf{P}\mathbf{W}_q, \mathbf{P}\mathbf{W}_k, \mathbf{P}\mathbf{W}_v,\\
    \operatorname{Attention}(\mathbf{Q, K, V}) &= \operatorname{SoftMax}\left(\mathbf{Q} \mathbf{K}^{T} / \sqrt{d}+\mathbf{B}\right) \mathbf{V},
\end{aligned}
\label{Eq.W-MSA}
\end{equation}
where $\mathbf{W}_q, \mathbf{W}_v, \mathbf{W}_v\in \mathbb{R}^{d\times d}$ are the \textit{query, key}, and \textit{value} projection matrices, respectively, $\mathbf{Q, K, V}$ denote the \textit{query, key} and \textit{value} results, respectively, and $\mathbf{B}\in {\mathbb{R}^{(2W^2-1)\times (2W^2-1)}}$ denotes the learnable relative position bias. The attention results are transformed into a feature map with the same size as $\mathbf{Z}$. It should be noted that we omit the multi-head format in Eq.~\eqref{Eq.W-MSA} for simplicity. Besides, we also use the shifted window partitioning scheme~\cite{liuSwinTransformerHierarchical2021} to capture dependencies across windows, and the W-MSA with shifted window partitioning is referred to SW-MSA.

\subsubsection{Multi-Channel Projection}

The projection operation reuses the measurement matrix to make the current image reconstruction more similar to the ground truth image, which can well utilize the prior knowledge of CS during the reconstruction process. It is necessary to integrate this CS characteristic into the transformer to achieve high performance. As shown in Eq.~\eqref{Eq_projection}, the projection operation updates each image block with a single channel. However, the features in a transformer architecture generally have smaller resolutions but much more channels than the single-channel image. Therefore, the projection operation cannot be directly applied to the transformer architecture.

To address this issue, we propose the multi-channel projection (MCP), which is derived from the original projection operation expressed in Eq.~\eqref{Eq_projection} but applies block-wise projection on the multi-channel feature domain. The details of MCP are shown in Fig.~\ref{Fig.mc-projection}. Specifically, MCP first reshapes the input feature into an image-like feature $\mathbf{X}'\in \mathbb{R}^{H\times W \times C'}$ using the pixel shuffle operation~\cite{PixelShuffle}, ensuring that it has the same resolution as the input image. Thereafter, MCP applies block-wise projection to each block of each channel of $\mathbf{X}'$ feature using the learnable measurement matrix $\mathbf{\Phi}$ and the adaptive sampling results $\mathbf{Y}$. The block-wise projection operation on each channel is calculated as follows:
\begin{equation}
    \begin{aligned}
        \mathbf{y}_{ij,c}' &= \mathbf{y}_{ij} - \mathbf{\Phi}[1:\mathbf{m}_{ij}, :] \mathfrak{T}(\mathbf{x}_{ij,c}')\\
    \mathfrak{T}(\mathbf{x}_{ij, c}'') &=  \mathfrak{T}(\mathbf{x}_{ij,c}') + \mathbf{\Phi}[1:\mathbf{m}_{ij}, :]^{T}\mathbf{y}_{ij,c}' / (1 + \mathbf{\alpha}_c),
    \end{aligned}
\end{equation}
where $c \in \{1, \cdots, C'\}$ is the channel index of $\mathbf{X}'$, $\mathfrak{T}(\mathbf{x}_{ij,c}')$ denotes the 1D vector form of each block $\mathbf{x}_{ij, c}'$ in the $c$-th channel of $\mathbf{X}'$, $\mathbf{m}_{ij}$ represents the measurement number of each block in the sampling process,  and updating step $\mathbf{\alpha}\in \mathbb{R}^{C'}$ is set to be learnable. By combining and reshaping the block-wise projection results, we can also obtain a feature $\mathbf{X}''$ with the same shape as $\mathbf{X}'$. Finally, MCP applies the pixel unshuffle operation to $\mathbf{X}''$ for restoring its channel number and original resolution.


\subsubsection{Projection-Based Transformer Block}

After constructing the MCP module for the transformer architecture, we develop a projection-based transformer block by integrating the MCP module into the transformer architecture. Fig.~\ref{Fig.mc-projection} shows the structure of the projection-based transformer block. Given the input feature $\mathbf{F}_{input}$, the entire process of the projection-based transformer block can be formulated as 
\begin{equation}
    \begin{aligned}
        \mathbf{Z}^{(0)} &= \mathbf{F}_{input}\\
        \mathbf{Z}^{(k)} & = \operatorname{WTB}(\mathbf{Z}^{(k-1)}), \quad k=1,\cdots,L\\
        \mathbf{F}_{output} &=\operatorname{MCP}(\operatorname{LN}(\mathbf{Z}^{(L)}))\\
    \end{aligned}
\end{equation}
where LN represents the layer normalization~\cite{ba2016layer}, and WTB indicates the following calculation process of an original transformer block: 
\begin{equation}
    \begin{aligned}
        \mathbf{T}^{(k)} & =\mathbf{Z}^{(k-1)}+ \operatorname{W/SW-MSA}(\operatorname{LN}(\mathbf{Z}^{(k-1)})) \\
        \mathbf{Z}^{(k)} & =\mathbf{T}^{(k)}+\operatorname{FFN}(\operatorname{LN}(\mathbf{T}^{(k)}))\\
    \end{aligned}
    \label{Eq.transformer_block}
\end{equation}
where FFN denotes a feed-forward network containing two fully connected layers. Following the shifted window partitioning scheme~\cite{liuSwinTransformerHierarchical2021}, the W-MSA is used when $k$ is odd and SW-MSA is used when $k$ is even. Because the MCP module reuses the learnable measurement matrix and only introduces a learnable updating step $\mathbf{\alpha}$, the projection-based transformer block can exploit the prior knowledge of CS with a small computation overhead and little extra parameters compared to  original stacked transformer blocks.

\subsubsection{Feature Fusion}
The encoder features are fused on the decoder side to enhance the feature representation. Given two features of the same size, they are directly concatenated in the channel dimension and a $1\times 1$  convolutional layer is used to reduce the number of channels in the concatenation result by $50\%$. Thereafter, a residual convolutional block is employed to further improve the feature fusion ability. Therefore, the feature fusion result has the same size as each input feature.

\subsection{Loss Function}
Following the settings of existing CS methods~\cite{CSNet+_2020,AMP-Net-2021}, we use the mean square error (MSE) to calculate the loss. The loss function of the proposed model comprises three parts $\mathcal{L}_1$, $\mathcal{L}_2$ and $\mathcal{L}_3$. 

Let $\{\mathbf{X}_k\}_{k=1}^{N}$ represent a training set. We first use the MSE to measure the difference between the input images and their corresponding reconstructed images as follows:
\begin{equation}
  \begin{split}
  \mathcal{L}_1 = \frac{1}{2N} \sum_{k = 1}^{N} \| \hat{\mathbf{X}}_k  -\mathbf{X}_k
  \|_2^2,
  \end{split}
\end{equation}
where $\hat{\mathbf{X}}_k$ denotes the reconstructed image of the $k$-th input image $\mathbf{X}_k$. Inspired by the cycle-consistent loss~\cite{zhu2017unpaired}, we also use the MSE to measure the difference between the measurements of the input images and the measurements of their corresponding reconstructed images as follows:
\begin{equation}
  \begin{split}
  \mathcal{L}_2 = \frac{1}{2N} \sum_{k = 1}^{N} \sum_{i = 1}^{h}\sum_{j = 1}^{w} \| y_{kij}  - \mathbf{\Phi}[1:\mathbf{m}_{kij}, :] \mathfrak{T}(\hat{\mathbf{x}}_{kij})
  \|_2^2,
  \end{split}
\end{equation}
where $\mathfrak{T}(\hat{\mathbf{x}}_{kij})$ denotes the 1D vector form of the image block $\hat{\mathbf{x}}_{kij}$ in the reconstructed image $\hat{\mathbf{X}}_k$. Considering that sparsity estimation is based on low-quality estimation of image, we further add an additional loss item to improve the accuracy of sparsity estimation:
\begin{equation}
  \mathcal{L}_3 = \frac{1}{2N} \sum_{k = 1}^{N} \| \hat{\mathbf{X}}_{k}^0  -\mathbf{X}_k
  \|_2^2
\end{equation}
where $\hat{\mathbf{X}}_{k}^0$ is the low-quality estimation of image $\mathbf{X}_k$.

Finally, we define the total loss of the proposed model as:
\begin{equation}
  \mathcal{L} = \mathcal{L}_1 + \lambda_1 \cdot  \mathcal{L}_2 + \lambda_2 \cdot\mathcal{L}_3
\end{equation}
where $\lambda_1$ and $\lambda_2$  are the regularization weights, which are set to 0.1 by default.

\section{Performance Evaluation}
\label{Sec.experiments}


\subsection{Experimental Settings}

\subsubsection{Datasets}

Compared to pure CNN models, a transformer model needs to be trained using a much larger training set~\cite{ViT}. To thoroughly evaluate the performance of the proposed model, we randomly select 40000 images from the COCO 2017 Unlabeled Images dataset\footnote[1]{\href{https://cocodataset.org}{https://cocodataset.org}}~\cite{COCO} as the training dataset and 100 images as the validation dataset. We conduct experiments to evaluate all the CS methods over five commonly used datasets: Set5~\cite{Set5-2012}, Set11~\cite{ReconNet_2016}, Set14~\cite{Set14-2012}, BSD100~\cite{MartinFTM01}, and Urban100~\cite{Huang-CVPR-2015}. 
For all used datasets, we only utilize the Y channel in the YCbCr color space of each color image.

\subsubsection{Implementation Details}

The image block size in the sampling process is set to $B=32$. Given the target sampling ratio $sr_t$, the initially sampling ratio $sr_{init}$ is set to $sr_t/2$ when $sr_t <= 0.1$ and to $sr_t/3$ when $sr_t>0.1$. The maximum sampling ratio $sr_{max}$ for any image block is $max(1.0, 2sr_t)$. The number of channels in the head module is set to $C_0=32$. From the top to the bottom levels of the Uformer, the number $L$ of the attention layers and FFN layers in the projection-based transformer block is set as $[4,4,6,6]$,  the number of heads in W/SW-MSA is set as $[1, 2, 4, 8]$, and the window sizes in W/SW-MSA are set to $[8, 8, 4, 4]$.

We train the proposed Uformer-ICS for a maximum of 100 epochs at each sampling ratio. Early-stopping is used to suspend training when there is no PSNR improvement on the validation set after the 50th epoch. The batch size is set to 16. For each input image in the training process, we first randomly flip and rotate it to augment the image features and then randomly crop the augmentation result into a sub-image of size $128\times 128$ for training. We use the Adam optimizer~\cite{Adam-2014} to optimize the parameters. The learning rate is initialized at 0.0001 and then decayed to half if there is no PSNR improvement on the validation dataset for five epochs in the training process. We use PyTorch framework to implement our models and conduct all the experiments on a computer with a RTX4090 GPU and an Intel i9-10920X CPU. It takes about 7 minutes to train one epoch. Our codes are available at Github\footnote[2]{\href{https://github.com/RedamancyAY/Uformer-ICS}{https://github.com/RedamancyAY/Uformer-ICS}}.

\begin{table}[!htbp]
    \centering
    \caption{Ablation studies on sparsity estimation methods. We list the PSNR results on Set11/Urban100 datasets at different sampling ratios ($sr$).}
      \vspace{-6pt}
    \label{Tab.ablation_study_adaptive}
    \renewcommand{\arraystretch}{1.1}
    \begin{tabular}{cccccc}
    \toprule
    \multirow{2}{*}{$sr$} & \multirow{2}{*}{\tabincell{c}{Non \\Adaptive}} & \multicolumn{3}{c}{Sparsity Estimation Methods} \\ \cmidrule(l){3-5}
                        & &  STD        & DIFF        & SM       \\ 
    \midrule
    0.01                & \response{\textbf{22.06}}/\response{21.86} & \response{21.95}/\response{21.92} & \response{22.01}/\response{\textbf{21.96}} & \response{22.04}/\response{21.94}           \\
    0.04                & \response{26.61}/\response{25.76} & \response{26.89}/\response{26.03} & \response{\textbf{27.03}}/\response{26.28} & \response{26.90}/\response{\textbf{26.29}}           \\
    0.1                 & \response{30.43}/\response{29.45} & \response{30.52}/\response{29.57} & \response{\textbf{30.73}}/\response{29.87} & \response{30.41}/\response{\textbf{30.00}}        \\
    0.25                & \response{34.48}/\response{33.93} & \response{34.24}/\response{34.17} & \response{\textbf{34.85}}/\response{34.55} & \response{34.55}/\response{\textbf{34.89}}         \\
    0.5                 & \response{38.35}/\response{38.45} & \response{38.54}/\response{39.20} & \response{39.10}/\response{39.69} & \response{\textbf{39.28}}/\response{\textbf{40.33}}         \\
    \midrule
    Avg. & \response{30.39}/\response{29.89} & \response{30.43}/\response{30.18} & \response{\textbf{30.74}}/\response{30.47} & \response{30.64}/\response{\textbf{30.69}} \\
    \bottomrule
    \end{tabular}
\end{table}

\subsubsection{Sampling and Reconstruction using One Model}
The proposed model can also sample and reconstruct images at arbitrary sampling ratios with only one-time training, which can significantly reduce the training time and storage burden. In our implementation, we term the proposed model with only one-time training for arbitrary sampling ratios as Uformer-ICS$^+$. In the training process of Uformer-ICS$^+$, each image in the training set is randomly assigned a target sampling ratio ranging from 0.01--0.5. Besides the training strategy, Uformer-ICS+ and Uformer-ICS have different sizes of the measurement matrices. The size of measurement matrix of Uformer-ICS is $(B^2\cdot sr_{max})\times B^2$, while that of Uformer-ICS$^+$ is $B^2\times B^2$ to deal with different sampling tasks at varying $sr_t$. We test the performances of both kinds of models in our experiments. 


\subsubsection{Competing Methods}
We compare our method with eight deep learning-based CS methods, which are developed using CNNs or Transformers and have shown state-of-the-art reconstruction performances.
\begin{itemize}
    \item SCSNet~\cite{SCSNet-2019}: This is a scalable sampling network that shares a single learnable measurement matrix for multiple reconstruction models.
    \item CSNet$^{+}$~\cite{CSNet+_2020}: This network utilizes several residual convolutional blocks for image reconstruction.
    \item ISTA-Net$^{++}$~\cite{you2021ISTA}: This scalable sampling network uses multiple non-learnable sampling matrices but only a single deep-unfolding multi-stage network. 
    \item AMP-Net~\cite{AMP-Net-2021}: This is a deep-unfolding multi-stage network, which unfolds the traditional AMP method~\cite{AMP-2009} and integrates deblocking modules to eliminate blocking artifacts.

    \item OCTUF~\cite{song2023optimization_OCTUF}: This work utilize optimization-inspired cross-attention transformer module to construct a lightweight unfolding framework for image CS.

    \item TransCS~\cite{shenTransCSTransformerBasedHybrid2022_TransCS}: This Transformer-CNN hybrid network contains a customized ISTA-based transformer backbone to model long-distance dependence and a auxiliary CNN to capture the local features.
    \item CSformer~\cite{CSformer-2023}: This hybrid network utilizes two stems to integrate the CNN and transformer architectures for improved representation learning. 
    \item DPC-DUN~\cite{Song2023-DPC-DUN}: This deep unfolding network introduces a path-controllable selector to dynamically select a rapid and appropriate route for each image.

 \end{itemize}
To keep consistency with other models, we set all the sampling matrices learnable for ISTA-Net$^{++}$~\cite{you2021ISTA} and DPC-DUN~\cite{Song2023-DPC-DUN}.
Additionally, four traditional CS methods, the TV~\cite{TV-2009}, BCS-FOCUSS~\cite{BCS-FOCUSS_2017}, DAMP~\cite{DAMP-2016}, and MH-BCS-SPL~\cite{MH-2011}, are also compared to demonstrate the powerful learning ability of the deep learning-based CS methods. We evaluate all of these methods using their publicly available implementation codes and retrain the deep-learning models on our constructed COCO-40000 training set.


\begin{table*}[!tbp]
    \centering
    \caption{Ablation study results of the proposed Uformer-ICS on Set11 and Urban100 datasets at different sampling ratios ($sr$). For the feature fusion, ``concat'' denotes the concatenation operation and ``add'' indicates the skip connection. }
      \vspace{-6pt}
    \label{Tab.ablation_study}
    \renewcommand{\arraystretch}{1.1}
    \begin{tabular}{cccccccccccccccc}
    \toprule
    \multirow{2}{*}{Setting} & \multirow{2}{*}{MCP} & \multirow{2}{*}{\tabincell{c}{Feature\\Fusion}} &\multirow{2}{*}{$\mathcal{L}_2$} & \multicolumn{5}{c}{Different $sr$ on Set11} & \multicolumn{5}{c}{Different $sr$ on Urban100} & \multirow{2}{*}{Parameters}\\ \cmidrule(lr){5-9} \cmidrule(lr){10-14}
     
     
     &  &  & & 0.01 & 0.04 & 0.1 & 0.25 & 0.5 & 0.01 & 0.04 & 0.1 & 0.25 & 0.5 & \\
     \midrule
    (a) & $\times$  & ``concat'' & \response{\Checkmark}  & \response{21.94}  &  \response{26.64}  &  \response{29.27}  &  \response{33.39}  &  \response{38.68}  &  \response{21.67}  &  \response{25.54}  &  \response{28.56}  &  \response{32.59}  &  \response{37.86} & 9.1588M\\
    (b) & \Checkmark   & ``add'' & \response{\Checkmark} & \response{21.97}  &  \response{\textbf{26.94}}  &  \response{30.68}  &  \response{34.54}  &  \response{\textbf{39.38}}  &  \response{21.94}  &  \response{26.27}  &  \response{29.99}  &  \response{34.86}  &  \response{40.07} & 9.1159M \\
    (c) & \Checkmark   & ``concat''& \response{$\times$}  &\response{21.99}  & \response{26.75}  & \response{\textbf{30.73}}  & \response{34.49}  & \response{38.57}  & \response{\textbf{21.96}}  &\response{ 26.23}  & \response{29.99}  & \response{34.68}  & \response{40.09} & 9.1589M\\
    (d) & \Checkmark   & ``concat''& \response{\Checkmark}  & \response{\textbf{22.04}}  &  \response{26.90}  &  \response{30.41}  &  \response{\textbf{34.55}}  &  \response{39.28}  &  \response{21.94}  &  \response{\textbf{26.29}}  &  \response{\textbf{30.00}}  &  \response{\textbf{34.89}}  &  \response{\textbf{40.33}} & 9.1589M\\
    \bottomrule
    \end{tabular}
\end{table*}

\begin{figure*}[!tb]
  \centering
    \includegraphics[width=0.8\linewidth]{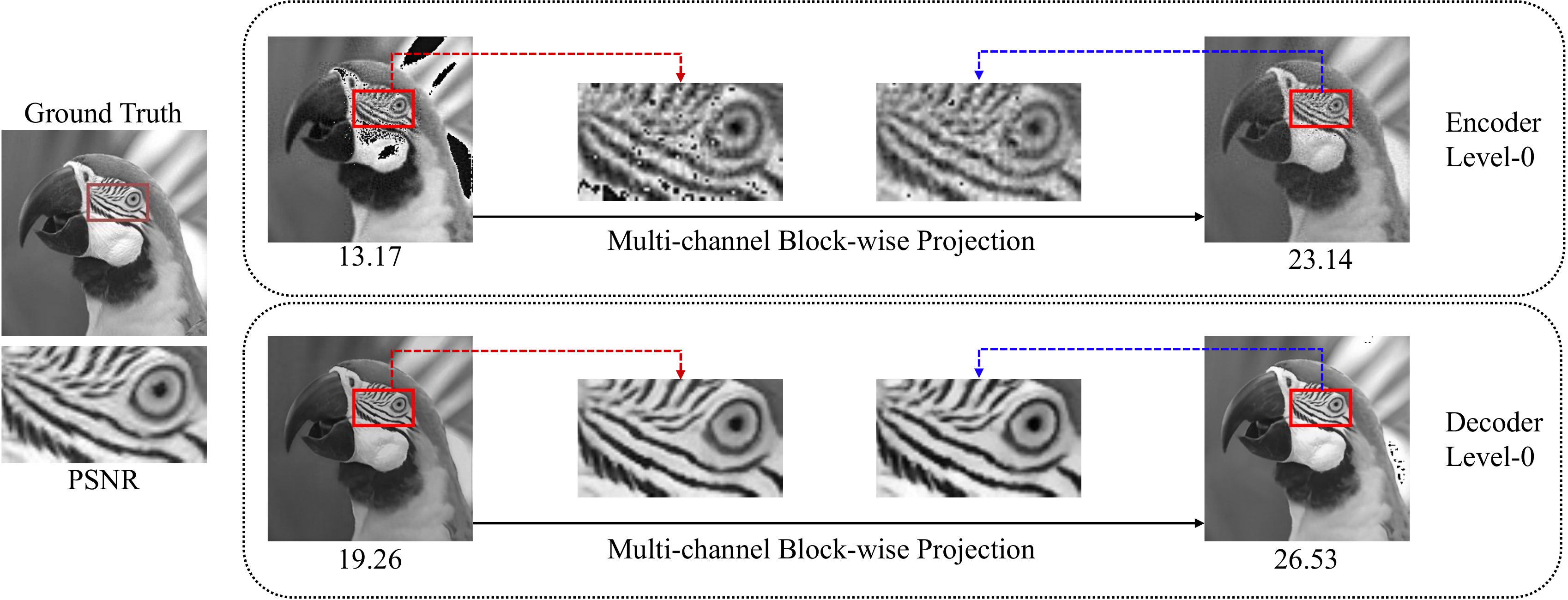}
      \vspace{-6pt}
    \caption{\response{Visual effects of the proposed multi-channel projection (MCP) modules at the sampling ratio of 0.1. It applies multi-channel block-wise projection to the input feature map for feature updating. Note that we only extract the image feature maps before and after the MCP modules for comparison, and} 
    we utilize the tail module $H_{tail}$ to convert these multi-channel feature maps into single-channel images for visualization.}
    \label{Fig.projection_res}
\end{figure*}

\subsection{Ablation Studies}
\label{Sec.discussion}
In this section, we conduct ablation studies on several components of the proposed Uformer-ICS to investigate their effectiveness and select the best settings.


\begin{table*}[!htbp]
  \centering
  \renewcommand{\arraystretch}{1.0}
  \caption{Quantitative performance comparisons of the proposed Uformer-ICS and Uformer-ICS$^+$ with existing CS methods under multiple sampling ratios ($sr$). The best and second best PSNR and SSIM scores are highlighted in \textcolor{red}{red} and \textcolor{blue}{blue}, respectively.}
  \vspace{-6pt}
  \label{Tab.Comparsion_CS}
  \setlength{\tabcolsep}{4mm}{
      \begin{tabular}{clcccccccccccc}
          \toprule
      \multirow{2}{*}{Dataset} & \multirow{2}{*}{Methods} & \multicolumn{5}{c}{$sr$ (PSNR/SSIM)}\\ \cmidrule(l){3-7}
      & & 0.01 & 0.04 & 0.1 & 0.25 & 0.5 \\ \midrule

\multirow{14}{*}{\tabincell{c}{Set5}} & TV~\cite{TV-2009} & 16.84/0.4412 & 21.34/0.5554 & 25.88/0.7239 & 30.96/0.8712 & 36.22/0.9465 \\
 & BCS-FOCUSS~\cite{BCS-FOCUSS_2017} & 16.79/0.4787 & 23.13/0.6549 & 27.30/0.7658 & 31.84/0.8693 & 36.26/0.9340 \\
 & BM3D-DAMP\cite{DAMP-2016} & 6.48/0.0080 & 14.70/0.4573 & 24.11/0.6849 & 31.24/0.8752 & 37.89/0.9461 \\
 & MH-BCS-SPL~\cite{MH-2011} & 14.59/0.3812 & 24.10/0.6709 & 28.66/0.8108 & 32.94/0.8975 & 36.99/0.9467 \\ 
 \cdashline{2-7}[1.5pt/5pt]
 & SCSNet~\cite{SCSNet-2019} & 22.58/0.5565 & 24.21/0.6552 & 28.65/0.8070 & 30.20/0.8862 & 34.79/0.9532 \\
 & ISTA-Net$^{++}$~\cite{you2021ISTA} & 24.17/0.6600 & 29.09/0.8297 & 33.09/0.9065 & 37.03/0.9502 & 40.90/0.9747\\
 & AMP-Net~\cite{AMP-Net-2021} & 24.65/0.6699 & 29.51/0.8394 & 33.63/0.9150 & 37.56/0.9564 & 41.93/0.9800\\
 & CSNet$^{+}$~\cite{CSNet+_2020} & 24.03/0.6461 & 28.72/0.8154 & 32.47/0.9013 & 36.17/0.9469 & 39.64/0.9735\\
 & OCTUF~\cite{song2023optimization_OCTUF} & \textcolor{blue}{24.89}/0.6901 & 30.05/0.8537 & 34.37/0.9234 & 38.37/0.9598 & 43.09/\textcolor{red}{0.9826}\\
 & TransCS~\cite{shenTransCSTransformerBasedHybrid2022_TransCS} & 24.32/0.6644 & 28.99/0.8235 & 33.29/0.9042 & 35.92/0.9351 & 39.39/0.9694\\
 & \response{DPC-DUN}~\cite{Song2023-DPC-DUN}& 24.33/0.6636 & 29.28/0.8397 & 33.51/0.9169 & 37.39/0.9576 & 41.79/0.9803\\
 & \response{CSformer}~\cite{CSformer-2023} &  24.65/\textcolor{blue}{0.6948} &  29.69/0.8520 & 33.33/0.9154  & 37.10/0.9551 & 40.97/0.9783 \\
 & \textbf{Uformer-ICS$^+$} & \textcolor{blue}{24.89}/0.6932 & \textcolor{blue}{30.40}/\textcolor{blue}{0.8572} &  \textcolor{blue}{34.61}/\textcolor{blue}{0.9238} & \textcolor{blue}{38.80}/\textcolor{blue}{0.9602} & \textcolor{blue}{43.68}/0.9811\\
 & \textbf{Uformer-ICS} & \textcolor{red}{25.15}/\textcolor{red}{0.7025} & \textcolor{red}{30.76}/\textcolor{red}{0.8637} & \textcolor{red}{34.97}/\textcolor{red}{0.9265} & \textcolor{red}{39.23}/\textcolor{red}{0.9612} & \textcolor{red}{44.43}/\textcolor{blue}{0.9824}\\
\midrule

\multirow{14}{*}{Set11} & TV~\cite{TV-2009}  & 15.49/0.3800 & 19.25/0.4988 & 22.78/0.6592 & 27.64/0.8296 & 32.81/0.9264 \\
 & BCS-FOCUSS~\cite{BCS-FOCUSS_2017} & 17.58/0.4488 & 20.96/0.5856 & 24.06/0.7075 & 28.41/0.8359 & 33.17/0.9203 \\
 & BM3D-DAMP\cite{DAMP-2016} & 5.18/0.0076 & 13.84/0.3904 & 21.21/0.6070 & 28.29/0.8484 & 35.19/0.9429 \\
 & MH-BCS-SPL~\cite{MH-2011} & 12.82/0.3205 & 21.49/0.6218 & 26.01/0.7862 & 30.53/0.8891 & 34.94/0.9447 \\ 
 \cdashline{2-7}[1.5pt/5pt]
 & ISTA-Net$^{++}$~\cite{you2021ISTA}& 21.23/0.5699 & 25.34/0.7640 & 28.83/0.8643 & 32.67/0.9316 & 36.65/0.9681 \\
 & SCSNet~\cite{SCSNet-2019} & 20.71/0.5478 & 22.42/0.6784 & 27.37/0.8454 & 28.36/0.8881 & 33.29/0.9542\\
 & AMP-Net~\cite{AMP-Net-2021} & 21.09/0.5734 & 25.53/0.7759 & 29.13/0.8740 & 33.04/0.9414 & 36.74/0.9735\\
 & CSNet$^{+}$~\cite{CSNet+_2020} & 20.98/0.5553 & 24.91/0.7503 & 28.40/0.8575 & 32.14/0.9252 & 33.79/0.9643\\
 & OCTUF~\cite{song2023optimization_OCTUF} & 21.74/0.5946 & \textcolor{blue}{26.52}/0.8001 & \textcolor{blue}{30.21}/0.8906 & 34.31/0.9499 & 38.13/\textcolor{blue}{0.9767}\\
 & TransCS~\cite{shenTransCSTransformerBasedHybrid2022_TransCS} & 21.30/0.5674 & 24.58/0.7541 & 28.42/0.8601 & 30.38/0.9102 & 33.53/0.9563\\
 & \response{DPC-DUN}~\cite{Song2023-DPC-DUN}& 21.13/0.5753 & 25.37/0.7778 & 29.05/0.8749 & 32.58/0.9422 & 36.89/0.9742\\
 & \response{CSformer}~\cite{CSformer-2023} & \textcolor{blue}{21.90}/\textcolor{blue}{0.6073} & 26.09/0.7989 & 29.02/0.8791 & 32.20/0.9430 & 35.40/0.9714\\
 & \textbf{Uformer-ICS$^+$} & 21.62/0.5993 & 26.36/\textcolor{blue}{0.8031} & 30.07/\textcolor{blue}{0.8911} & \textcolor{blue}{34.37}/\textcolor{blue}{0.9486} & \textcolor{blue}{38.55}/0.9740\\
 & \textbf{Uformer-ICS} & \textcolor{red}{22.04}/\textcolor{red}{0.6130} & \textcolor{red}{26.90}/\textcolor{red}{0.8173} & \textcolor{red}{30.41}/\textcolor{red}{0.9003} & \textcolor{red}{34.55}/\textcolor{red}{0.9522} & \textcolor{red}{39.28}/\textcolor{red}{0.9768}\\
\midrule

\multirow{14}{*}{Set14} & TV~\cite{TV-2009}  & 17.01/0.4025 & 20.82/0.5008 & 24.32/0.6456 & 28.48/0.7988 & 33.35/0.9071 \\
 & BCS-FOCUSS~\cite{BCS-FOCUSS_2017} & 18.32/0.4586 & 21.99/0.5716 & 25.26/0.6816 & 29.01/0.8082 & 33.19/0.9023 \\
 & BM3D-DAMP\cite{DAMP-2016} & 6.17/0.0076 & 15.08/0.4032 & 22.73/0.5794 & 28.18/0.7630 & 34.29/0.8901 \\
 & MH-BCS-SPL~\cite{MH-2011} & 14.36/0.3383 & 22.65/0.5878 & 26.63/0.7308 & 30.46/0.8489 & 34.49/0.9243 \\\cdashline{2-7}[1.5pt/5pt]
 & ISTA-Net$^{++}$~\cite{you2021ISTA}& 23.03/0.5770 & 26.54/0.7114 & 29.49/0.8153 & 33.20/0.9022 & 37.23/0.9549\\
 & SCSNet~\cite{SCSNet-2019} & 22.58/0.5565 & 24.21/0.6552 & 28.65/0.8070 & 30.20/0.8862 & 34.79/0.9532 \\
 & AMP-Net~\cite{AMP-Net-2021} & 23.23/0.5804 & 26.83/0.7202 & 29.95/0.8277 & 33.93/0.9158 & 38.51/0.9642\\
 & CSNet$^{+}$~\cite{CSNet+_2020} & 22.74/0.5619 & 26.09/0.7012 & 29.02/0.8117 & 32.52/0.9009 & 36.20/0.9556\\
 & OCTUF~\cite{song2023optimization_OCTUF} & 23.45/0.5909 & 27.29/0.7333 & 30.52/0.8392 & 34.76/\textcolor{blue}{0.9237} & 39.64/\textcolor{red}{0.9684}\\
 & TransCS~\cite{shenTransCSTransformerBasedHybrid2022_TransCS} & 23.03/0.5708 & 26.24/0.7029 & 29.25/0.8174 & 31.83/0.8934 & 36.04/0.9541\\
 &\response{DPC-DUN}~\cite{Song2023-DPC-DUN}& 22.95/0.5769 & 26.64/0.7206 & 29.79/0.8269 & 33.72/0.9174 & 38.34/0.9648\\
 & \response{CSformer}~\cite{CSformer-2023} & 23.52/0.5954 & 27.10/0.7357 & 29.75/0.8327 & 33.49/0.9187 & 37.36/0.9635\\
 & \textbf{Uformer-ICS$^+$} & \textcolor{blue}{23.73}/\textcolor{blue}{0.5987} & \textcolor{blue}{27.58}/\textcolor{blue}{0.7381} & \textcolor{blue}{30.89}/\textcolor{blue}{0.8405} & \textcolor{blue}{35.15}/0.9218 & \textcolor{blue}{39.68}/0.9645\\
 & \textbf{Uformer-ICS} & \textcolor{red}{24.00}/\textcolor{red}{0.6059} & \textcolor{red}{27.95}/\textcolor{red}{0.7474} & \textcolor{red}{31.28}/\textcolor{red}{0.8461} & \textcolor{red}{35.68}/\textcolor{red}{0.9253} & \textcolor{red}{40.48}/\textcolor{blue}{0.9669}\\
\midrule

\multirow{14}{*}{BSD100} & TV~\cite{TV-2009}  & 17.63/0.4174 & 21.26/0.5054 & 24.68/0.6249 & 27.93/0.7719 & 31.97/0.8920 \\
 & BCS-FOCUSS~\cite{BCS-FOCUSS_2017} & 19.10/0.4587 & 22.89/0.5625 & 25.25/0.6566 & 28.22/0.7795 & 31.86/0.8834 \\
 & BM3D-DAMP\cite{DAMP-2016} & 6.79/0.0065 & 15.84/0.4096 & 23.06/0.5513 & 26.97/0.6979 & 31.02/0.8265 \\
 & MH-BCS-SPL~\cite{MH-2011} & 15.38/0.3614 & 23.15/0.5645 & 25.72/0.6754 & 28.98/0.8049 & 32.65/0.9004 \\ \cdashline{2-7}[1.5pt/5pt]
 & ISTA-Net$^{++}$~\cite{you2021ISTA} & 23.89/0.5540 & 26.29/0.6697 & 28.58/0.7763 & 31.80/0.8798 & 35.84/0.9481 \\
 & SCSNet~\cite{SCSNet-2019} & 23.64/0.5424 & 25.44/0.6382 & 28.30/0.7732 & 30.64/0.8743 & 35.22/0.9535 \\
 & CSNet$^+$~\cite{CSNet+_2020} & 23.70/0.5456 &	26.15/0.6643 &	28.39/0.7752 &	31.49/0.8810 &	35.49/0.9530 \\
 & AMP-Net~\cite{AMP-Net-2021} & 23.93/0.5544 &	26.46/0.6765&	28.88/0.7882&	32.38/0.8952&	36.90/0.9601 \\
 & OCTUF~\cite{song2023optimization_OCTUF} & 24.09/0.5627	& 26.71/0.6870	& 29.23/0.7976 &	32.83/0.9001 &	37.72/\textcolor{red}{0.9647} \\
 & TransCS~\cite{shenTransCSTransformerBasedHybrid2022_TransCS} & 23.90/0.5494 &	26.24/0.6648 &	28.70/0.7803 &	31.50/0.8768 &	35.63/0.9511 \\
 &\response{DPC-DUN}~\cite{Song2023-DPC-DUN}& 23.82/0.5538 & 26.41/0.6779 & 28.83/0.7895 & 32.39/0.8987 & 36.89/0.9613\\
 & \response{CSformer}~\cite{CSformer-2023} & 23.95/0.5642 & 26.60/0.6909 & 28.79/0.7923 & 32.17/0.8975 & 36.44/0.9613\\
 & \textbf{Uformer-ICS$^{+}$} &\textcolor{blue}{24.27}/\textcolor{blue}{0.5660} & \textcolor{blue}{27.13}/\textcolor{blue}{0.6948} & \textcolor{blue}{29.74}/\textcolor{blue}{0.8041} & \textcolor{blue}{33.67}/\textcolor{blue}{0.9031} & \textcolor{blue}{38.74}/0.9623 \\
  & \textbf{Uformer-ICS} & \textcolor{red}{24.39}/\textcolor{red}{0.5697} & \textcolor{red}{27.30}/\textcolor{red}{0.7013} & \textcolor{red}{29.89}/\textcolor{red}{0.8065} & \textcolor{red}{33.90}/\textcolor{red}{0.9049} & \textcolor{red}{39.26}/\textcolor{red}{0.9647} \\
\midrule

\multirow{14}{*}{Urban100} & TV~\cite{TV-2009}  & 16.17/0.3819 & 19.31/0.4766 & 22.07/0.6181 & 26.05/0.7889 & 31.06/0.9107 \\
 & BCS-FOCUSS~\cite{BCS-FOCUSS_2017} & 16.68/0.4116 & 20.17/0.5279 & 22.89/0.6435 & 26.65/0.7864 & 31.19/0.8947 \\
 & BM3D-DAMP\cite{DAMP-2016} & 6.13/0.0065 & 14.36/0.3709 & 21.03/0.5650 & 27.05/0.7947 & 33.55/0.9194 \\
 & MH-BCS-SPL~\cite{MH-2011} & 13.63/0.3177 & 20.63/0.5464 & 24.95/0.7399 & 29.63/0.8720 & 33.92/0.9380 \\ \cdashline{2-7}[1.5pt/5pt]
 & ISTA-Net$^{++}$~\cite{you2021ISTA} & 20.89/0.5325 & 23.93/0.6871 & 26.99/0.8096 & 31.16/0.9067 & 35.81/0.9609 \\
 & SCSNet~\cite{SCSNet-2019} & 20.59/0.5123 & 22.25/0.6109 & 26.10/0.7887 & 27.53/0.8547 & 31.98/0.9383 \\
 & AMP-Net~\cite{AMP-Net-2021} & 21.02/0.5367 & 24.27/0.7021 & 27.65/0.8316 & 32.22/0.9244 & 37.26/0.9704 \\
 & CSNet$^{+}$~\cite{CSNet+_2020} & 20.69/0.5183 & 23.61/0.6729 & 26.41/0.7993 & 30.15/0.8988 & 34.08/0.9560\\
 & OCTUF~\cite{song2023optimization_OCTUF} & 21.41/0.5582 & 25.22/0.7413 & 28.96/\textcolor{blue}{0.8608} & 33.68/\textcolor{blue}{0.9374} & 38.85/\textcolor{red}{0.9767}\\
 & TransCS~\cite{shenTransCSTransformerBasedHybrid2022_TransCS} & 21.00/0.5288 & 23.74/0.6747 & 27.17/0.8175 & 30.03/0.8917 & 34.71/0.9572\\
 &\response{DPC-DUN}~\cite{Song2023-DPC-DUN} & 20.91/0.5371 & 24.20/0.7057 & 27.44/0.8295 & 32.02/0.9262 & 36.99/0.9707\\
 & \response{CSformer}~\cite{CSformer-2023} & \textcolor{blue}{21.60}/\textcolor{blue}{0.5727} & 25.17/\textcolor{blue}{0.7478} & 27.96/0.8455 & 31.81/0.9271 & 35.36/0.9663\\
 & \textbf{Uformer-ICS$^+$} & 21.58/0.5649 & \textcolor{blue}{25.41}/0.7459 & \textcolor{blue}{29.05}/0.8573 & \textcolor{blue}{33.83}/0.9349 & \textcolor{blue}{39.00}/0.9719\\
 & \textbf{Uformer-ICS} & \textcolor{red}{21.94}/\textcolor{red}{0.5828} & \textcolor{red}{26.29}/\textcolor{red}{0.7750} & \textcolor{red}{30.00}/\textcolor{red}{0.8765} & \textcolor{red}{34.89}/\textcolor{red}{0.9421} & \textcolor{red}{40.33}/\textcolor{blue}{0.9760}\\
      \bottomrule
      \end{tabular}
  }
\end{table*}

\begin{figure*}[!htbp]
    \centering
    \begin{minipage}[b]{1.0\linewidth}
        \centering
        \begin{minipage}[b]{0.12\linewidth}
          \centering
            \centerline{\scriptsize{Ground Truth}}
            \vspace{2pt}
            \centerline{\includegraphics[width=1\linewidth]{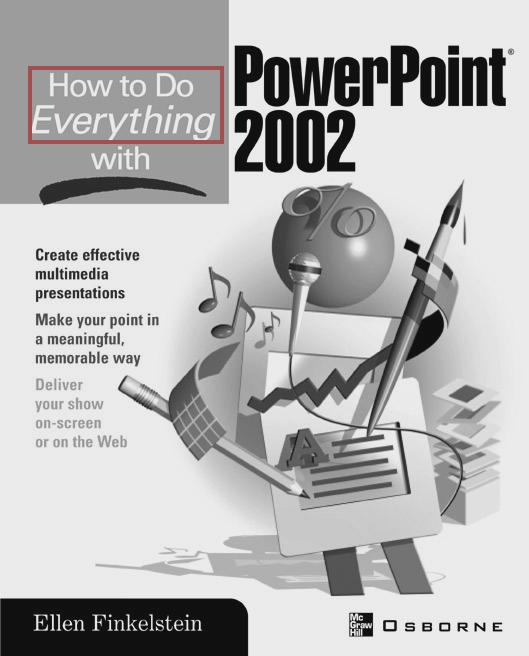}}
            \vspace{2pt}
            \centerline{\includegraphics[width=1\linewidth]{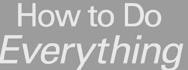}}
            \centerline{\scriptsize{PSNR/SSIM}}
        \end{minipage}
        \begin{minipage}[b]{0.12\linewidth}
            \centerline{\scriptsize{TV~\cite{TV-2009}}}
            \vspace{2pt}
            \centerline{\includegraphics[width=1\linewidth]{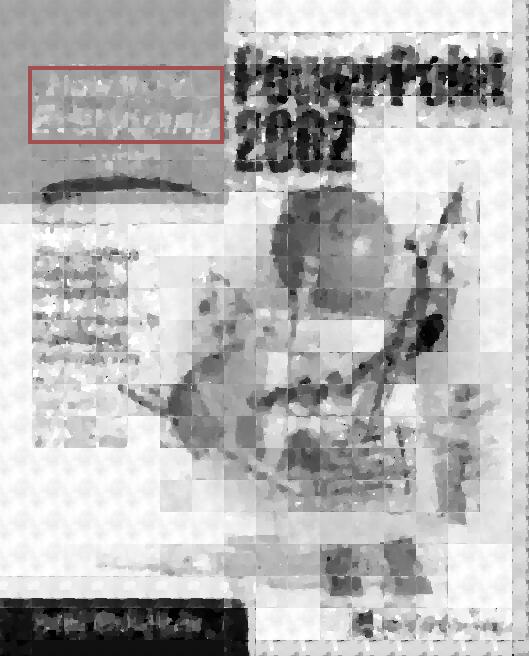}}
            \vspace{2pt}
            \centerline{\includegraphics[width=1\linewidth]{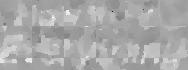}}
            \centerline{\scriptsize{17.94/0.5945}}
        \end{minipage}
        \begin{minipage}[b]{0.12\linewidth}
            \centerline{\scriptsize{BCS-FOCUSS~\cite{BCS-FOCUSS_2017}}}
            \vspace{2pt}
            \centerline{\includegraphics[width=1\linewidth]{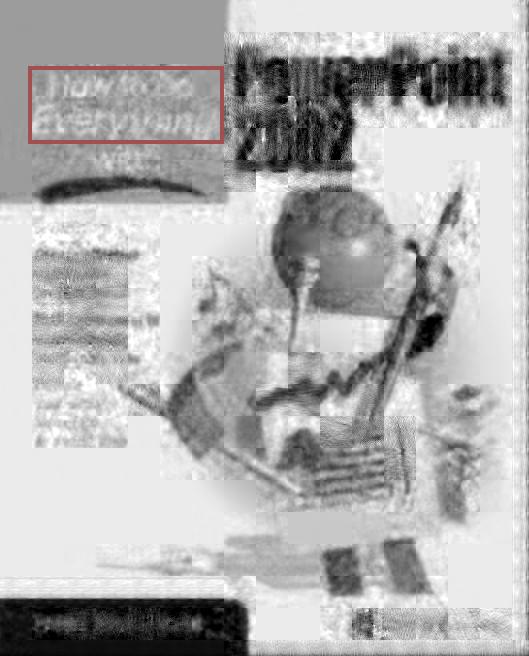}}
            \vspace{2pt}
            \centerline{\includegraphics[width=1\linewidth]{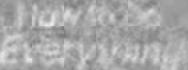}}
            \centerline{\scriptsize{18.15/0.6159}}
        \end{minipage}
        \begin{minipage}[b]{0.12\linewidth}
            \centerline{\scriptsize{BM3D-DAMP~\cite{DAMP-2016}}}
            \vspace{2pt}
            \centerline{\includegraphics[width=1\linewidth]{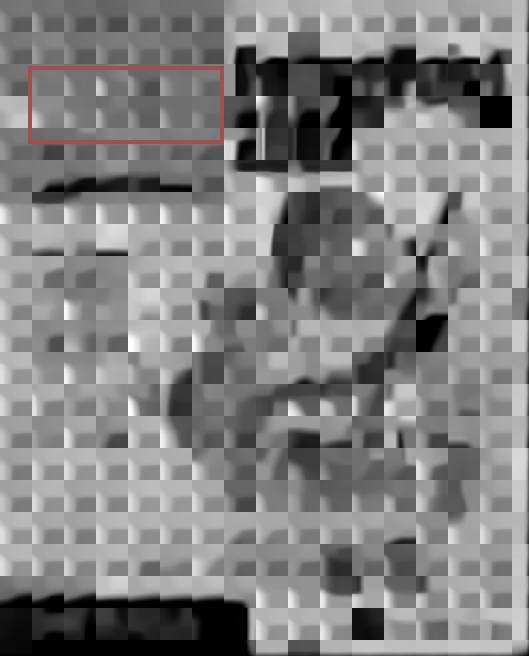}}
            \vspace{2pt}
            \centerline{\includegraphics[width=1\linewidth]{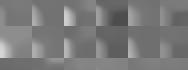}}
            \centerline{\scriptsize{11.21/0.4367}}
        \end{minipage}
        \begin{minipage}[b]{0.12\linewidth}
            \centerline{\scriptsize{MH-BCS-SPL~\cite{MH-2011}}}
            \vspace{2pt}
            \centerline{\includegraphics[width=1\linewidth]{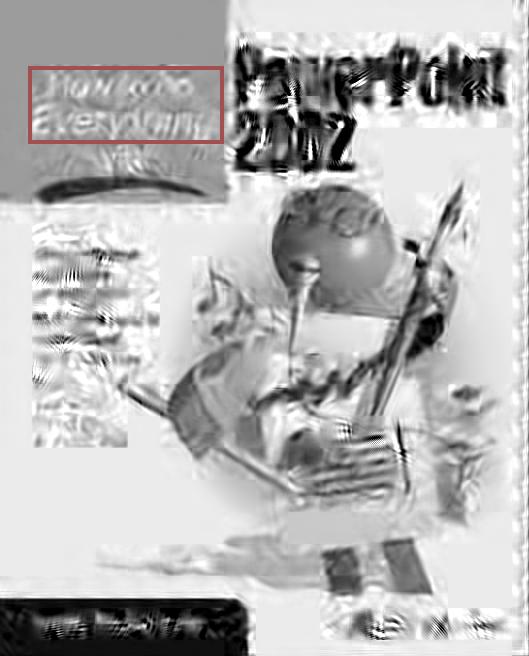}}
            \vspace{2pt}
            \centerline{\includegraphics[width=1\linewidth]{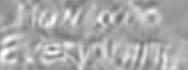}}
            \centerline{\scriptsize{19.08/0.6740}}
        \end{minipage}
        \begin{minipage}[b]{0.12\linewidth}
            \centerline{\scriptsize{CSNet$^+$~\cite{CSNet+_2020}}}
            \vspace{2pt}
            \centerline{\includegraphics[width=1\linewidth]{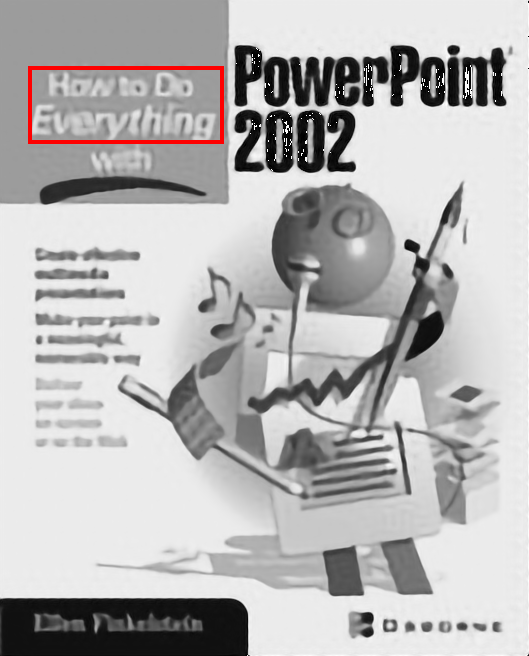}}
            \vspace{2pt}
            \centerline{\includegraphics[width=1\linewidth]{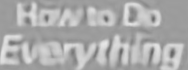}}
            \centerline{\scriptsize{22.69/0.8654}}
        \end{minipage}
        \begin{minipage}[b]{0.12\linewidth}
            \centerline{\scriptsize{AMP-Net~\cite{AMP-Net-2021}}}
            \vspace{2pt}
            \centerline{\includegraphics[width=1\linewidth]{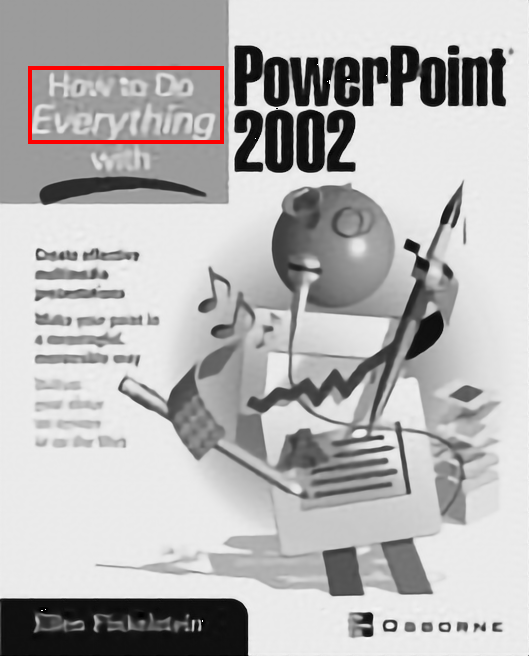}}
            \vspace{2pt}
            \centerline{\includegraphics[width=1\linewidth]{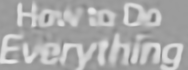}}
            \centerline{\scriptsize{25.56/0.9082}}
        \end{minipage}
        \begin{minipage}[b]{0.12\linewidth}
            \centerline{\scriptsize{ISTA-Net$^+$$^+$~\cite{you2021ISTA}}}
            \vspace{2pt}
            \centerline{\includegraphics[width=1\linewidth]{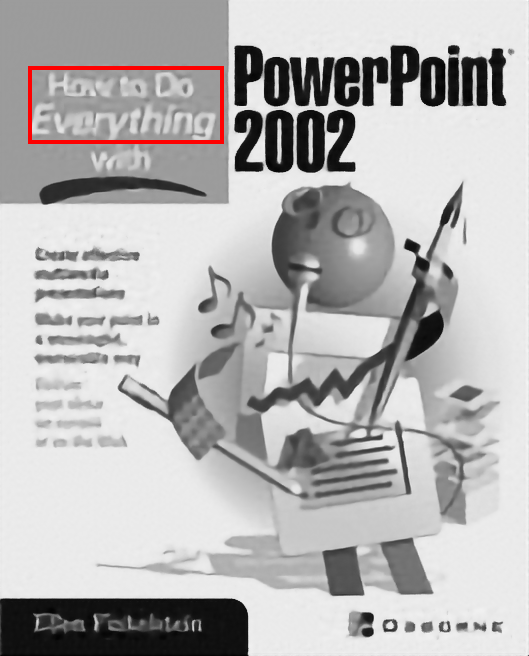}}
            \vspace{2pt}
            \centerline{\includegraphics[width=1\linewidth]{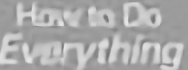}}
            \centerline{\scriptsize{25.14/0.8963}}
        \end{minipage}
        
        \vspace{10pt}

        \begin{minipage}[b]{0.12\linewidth}
            \centerline{\scriptsize{SCSNet$^+$~\cite{SCSNet-2019}}}
            \vspace{2pt}
            \centerline{\includegraphics[width=1\linewidth]{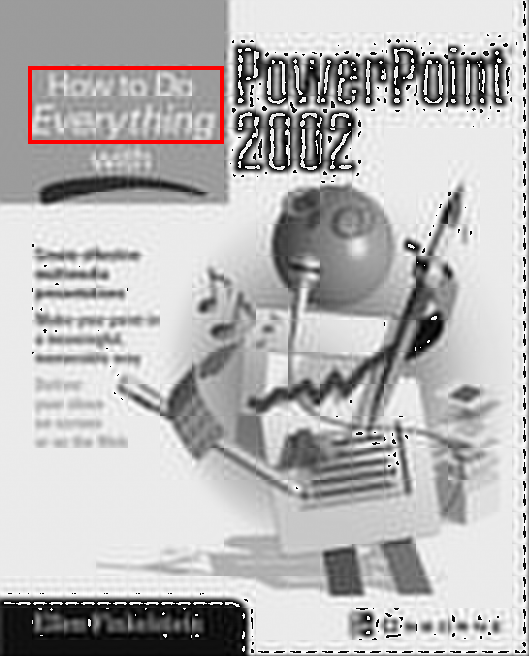}}
            \vspace{2pt}
            \centerline{\includegraphics[width=1\linewidth]{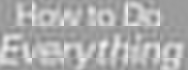}}
            \centerline{\scriptsize{16.55/0.6970}}
        \end{minipage}
        \begin{minipage}[b]{0.12\linewidth}
            \centerline{\scriptsize{OCTUF~\cite{song2023optimization_OCTUF}}}
            \vspace{2pt}
            \centerline{\includegraphics[width=1\linewidth]{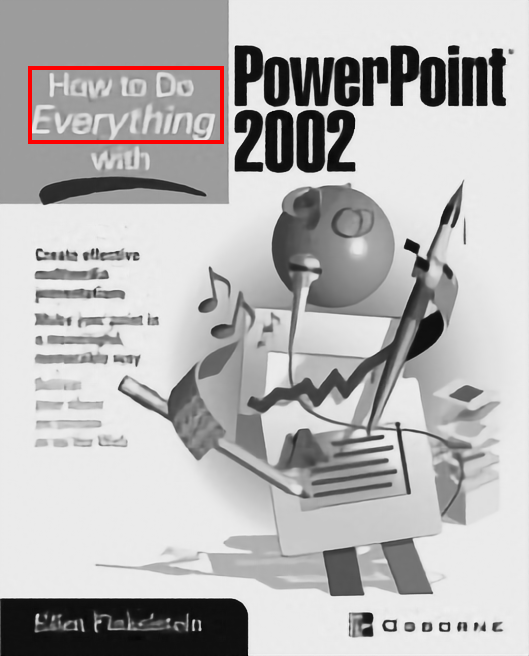}}
            \vspace{2pt}
            \centerline{\includegraphics[width=1\linewidth]{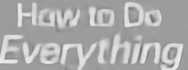}}
            \centerline{\scriptsize{26.41/0.9305}}
        \end{minipage}
        \begin{minipage}[b]{0.12\linewidth}
            \centerline{\scriptsize{TransCS~\cite{shenTransCSTransformerBasedHybrid2022_TransCS}}}
            \vspace{2pt}
            \centerline{\includegraphics[width=1\linewidth]{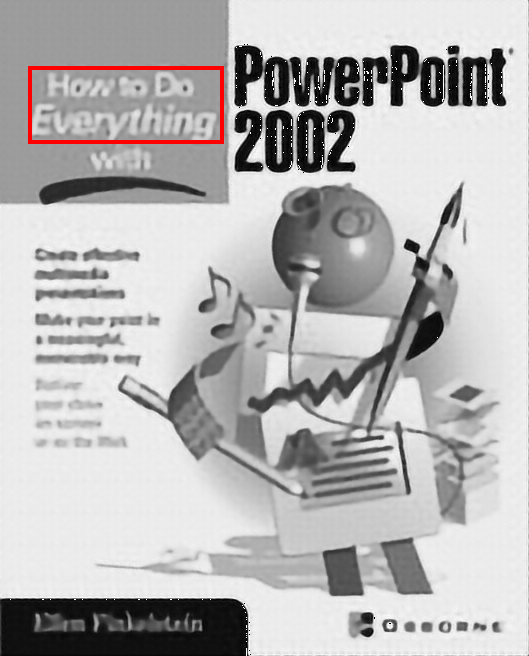}}
            \vspace{2pt}
            \centerline{\includegraphics[width=1\linewidth]{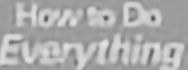}}
            \centerline{\scriptsize{23.05/0.8538}}
        \end{minipage}
        \begin{minipage}[b]{0.12\linewidth}
            \centerline{\scriptsize{\response{CSformer}~\cite{CSformer-2023}}}
            \vspace{2pt}
            \centerline{\includegraphics[width=1\linewidth]{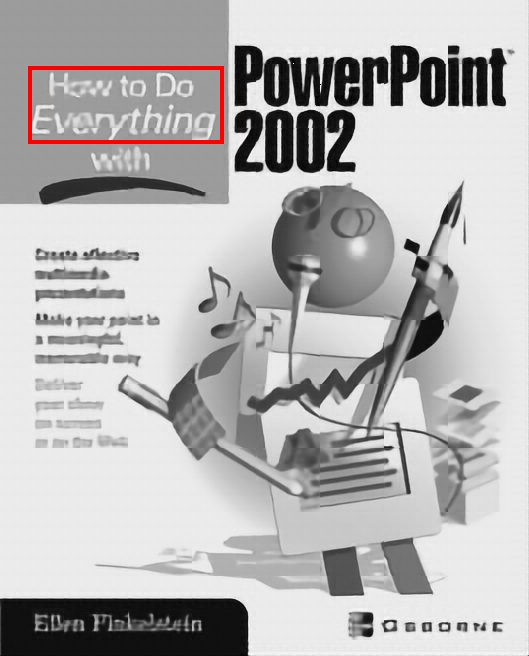}}
            \vspace{2pt}
            \centerline{\includegraphics[width=1\linewidth]{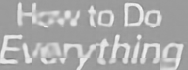}}
            \centerline{\scriptsize{\response{25.99/0.9216}}}
        \end{minipage}
        \begin{minipage}[b]{0.12\linewidth}
            \centerline{\scriptsize{\response{DPC-DUN}~\cite{Song2023-DPC-DUN}}}
            \vspace{2pt}
            \centerline{\includegraphics[width=1\linewidth]{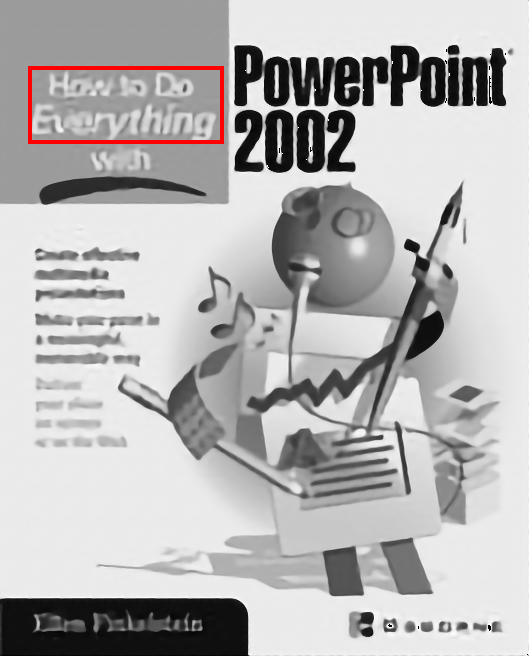}}
            \vspace{2pt}
            \centerline{\includegraphics[width=1\linewidth]{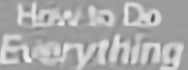}}
            \centerline{\scriptsize{\response{24.41/0.8967}}}
        \end{minipage}
        \begin{minipage}[b]{0.12\linewidth}
            \centerline{\scriptsize{\response{Ours-Non-Adaptive}}}
            \vspace{2pt}
            \centerline{\includegraphics[width=1\linewidth]{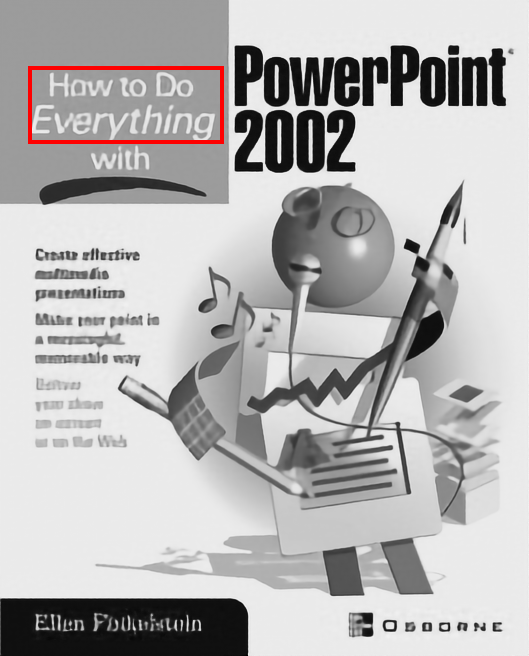}}
            \vspace{2pt}
            \centerline{\includegraphics[width=1\linewidth]{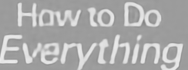}}
            \centerline{\scriptsize{\response{27.48/0.9473}}}
        \end{minipage}
        \begin{minipage}[b]{0.12\linewidth}
            \centerline{\scriptsize{\textbf{Uformer-ICS}$^+$}}
            \vspace{2pt}
            \centerline{\includegraphics[width=1\linewidth]{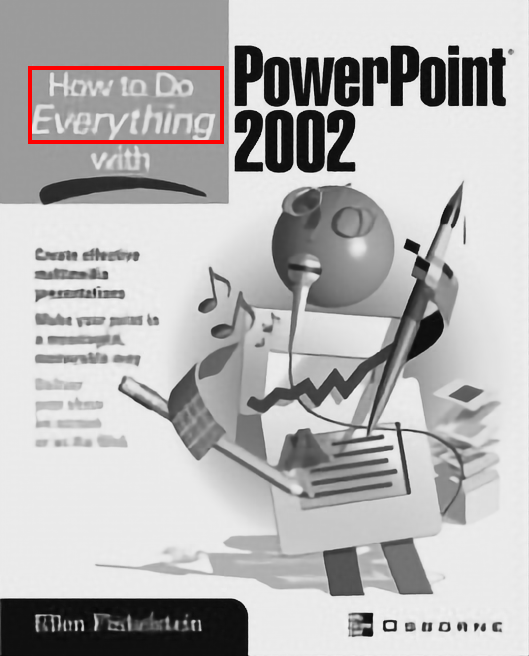}}
            \vspace{2pt}
            \centerline{\includegraphics[width=1\linewidth]{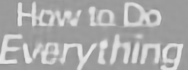}}
            \centerline{\scriptsize{\textcolor{blue}{26.72/0.9325}}}
        \end{minipage}
        \begin{minipage}[b]{0.12\linewidth}
            \centerline{\scriptsize{\textbf{Uformer-ICS}}}
            \vspace{2pt}
            \centerline{\includegraphics[width=1\linewidth]{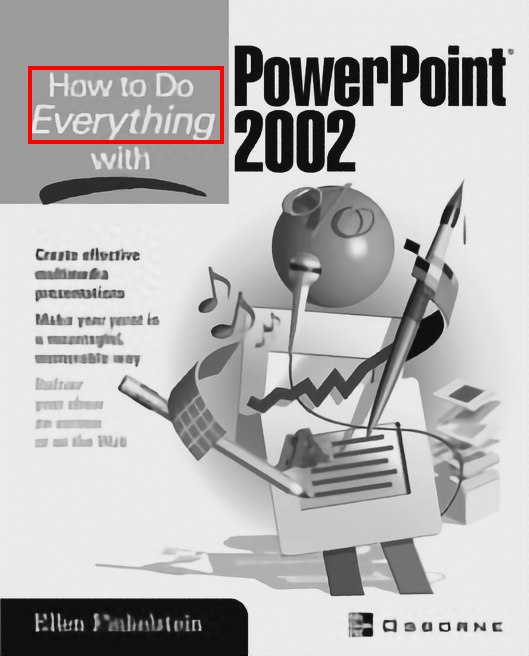}}
            \vspace{2pt}
            \centerline{\includegraphics[width=1\linewidth]{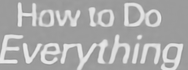}}
            \centerline{\scriptsize{\textbf{\textcolor{blue}{27.56/0.9510}}}}
        \end{minipage}
    \end{minipage}
            
    \centerline{\scriptsize{(a) Reconstruction results of the image ``ppt3'' at $sr$ = 0.04}}    
    \vspace{10pt}

    \begin{minipage}[b]{1.0\linewidth}
        \centering
        \begin{minipage}[b]{0.12\linewidth}
            \centerline{\scriptsize{Ground Truth}}
            \vspace{2pt}
            \centerline{\includegraphics[width=1\linewidth]{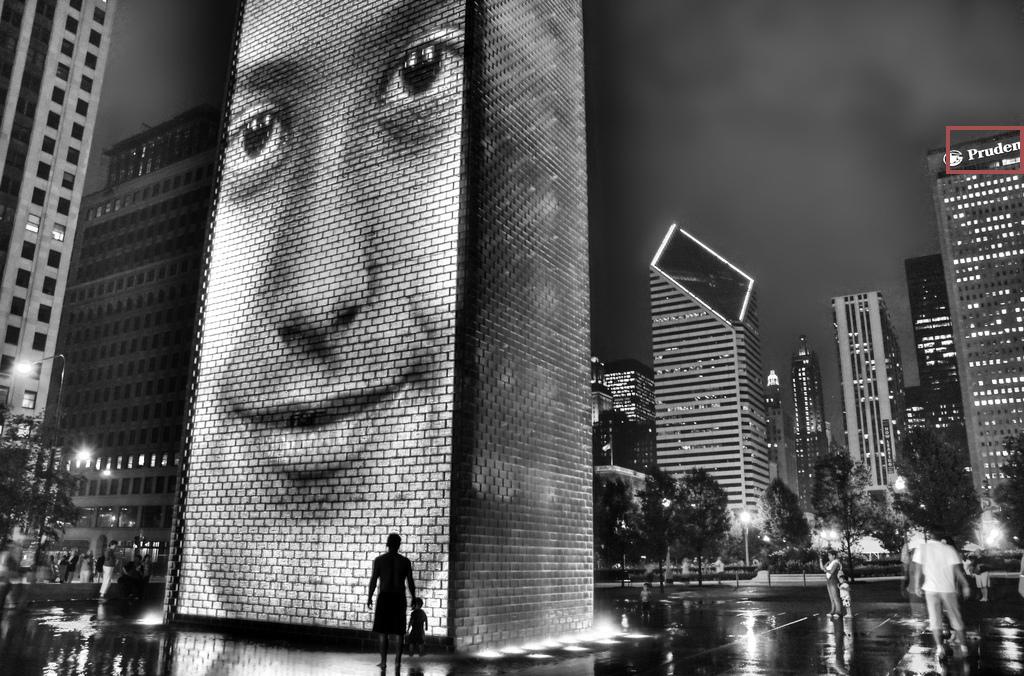}}
            \vspace{2pt}
            \centerline{\includegraphics[width=1\linewidth]{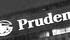}}
            \centerline{\scriptsize{PSNR/SSIM}}
        \end{minipage}
        \begin{minipage}[b]{0.12\linewidth}
            \centerline{\scriptsize{TV~\cite{TV-2009}}}
            \vspace{2pt}
            \centerline{\includegraphics[width=1\linewidth]{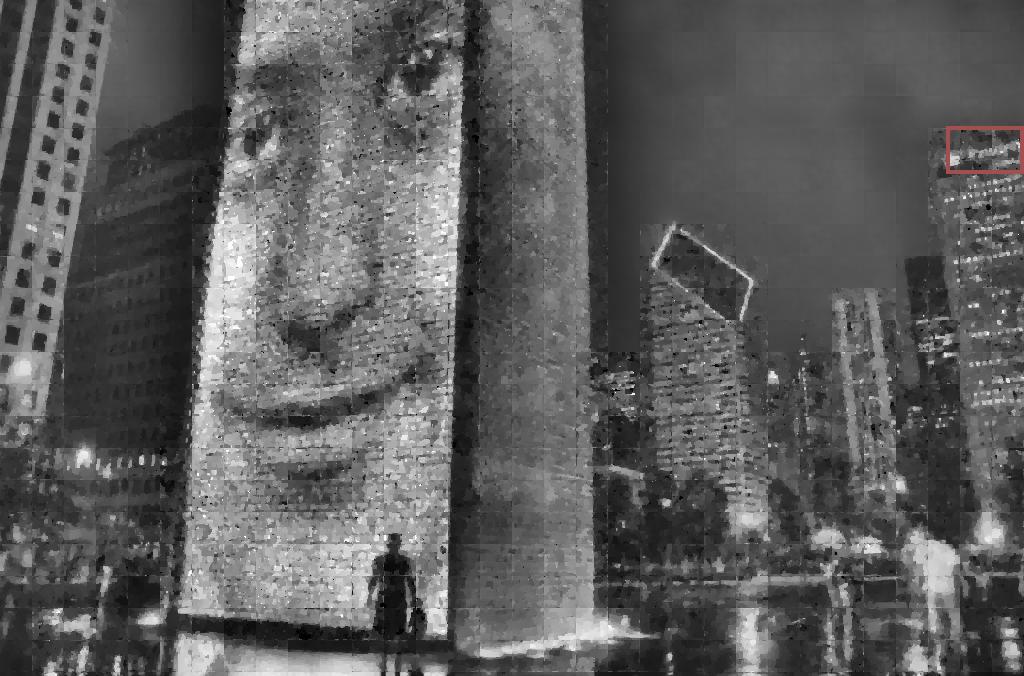}}
            \vspace{2pt}
            \centerline{\includegraphics[width=1\linewidth]{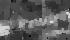}}
            \centerline{\scriptsize{20.70/0.5997}}
        \end{minipage}
        \begin{minipage}[b]{0.12\linewidth}
            \centerline{\scriptsize{BCS-FOCUSS~\cite{BCS-FOCUSS_2017}}}
            \vspace{2pt}
            \centerline{\includegraphics[width=1\linewidth]{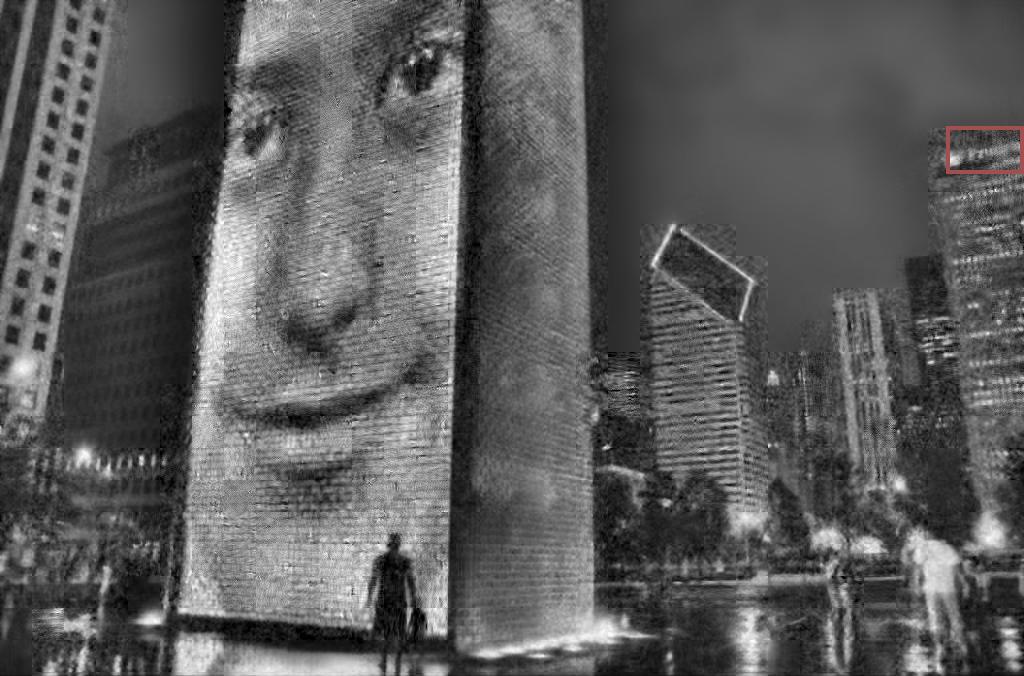}}
            \vspace{2pt}
            \centerline{\includegraphics[width=1\linewidth]{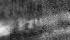}}
            \centerline{\scriptsize{21.52/0.6422}}
        \end{minipage}
        \begin{minipage}[b]{0.12\linewidth}
            \centerline{\scriptsize{BM3D-DAMP~\cite{DAMP-2016}}}
            \vspace{2pt}
            \centerline{\includegraphics[width=1\linewidth]{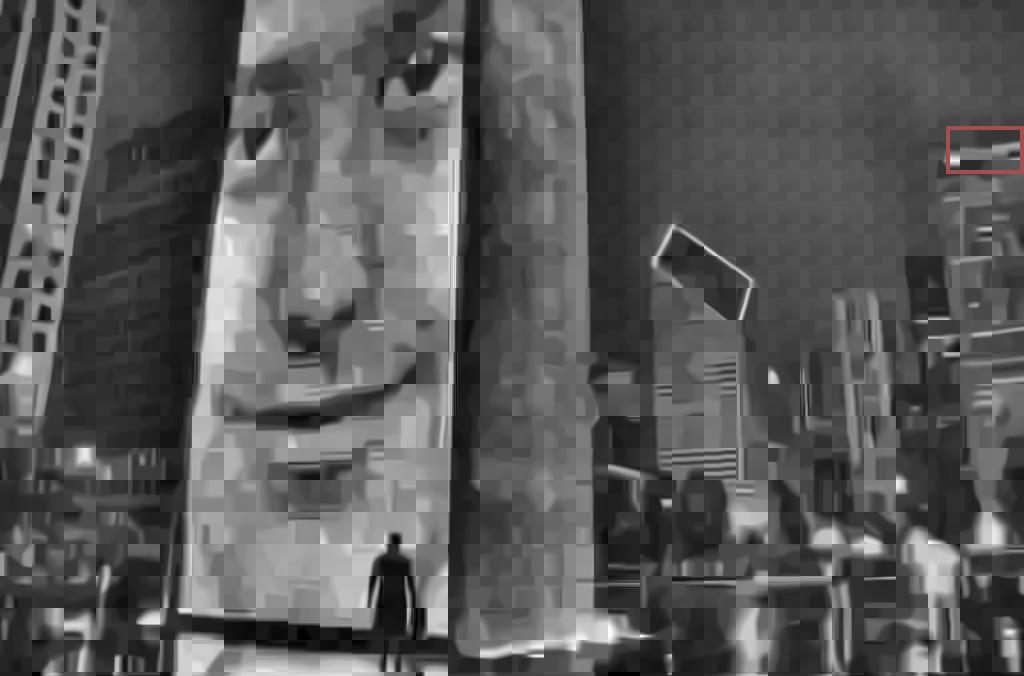}}
            \vspace{2pt}
            \centerline{\includegraphics[width=1\linewidth]{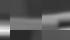}}
            \centerline{\scriptsize{19.85/0.5164}}
        \end{minipage}
        \begin{minipage}[b]{0.12\linewidth}
            \centerline{\scriptsize{MH-BCS-SPL~\cite{MH-2011}}}
            \vspace{2pt}
            \centerline{\includegraphics[width=1\linewidth]{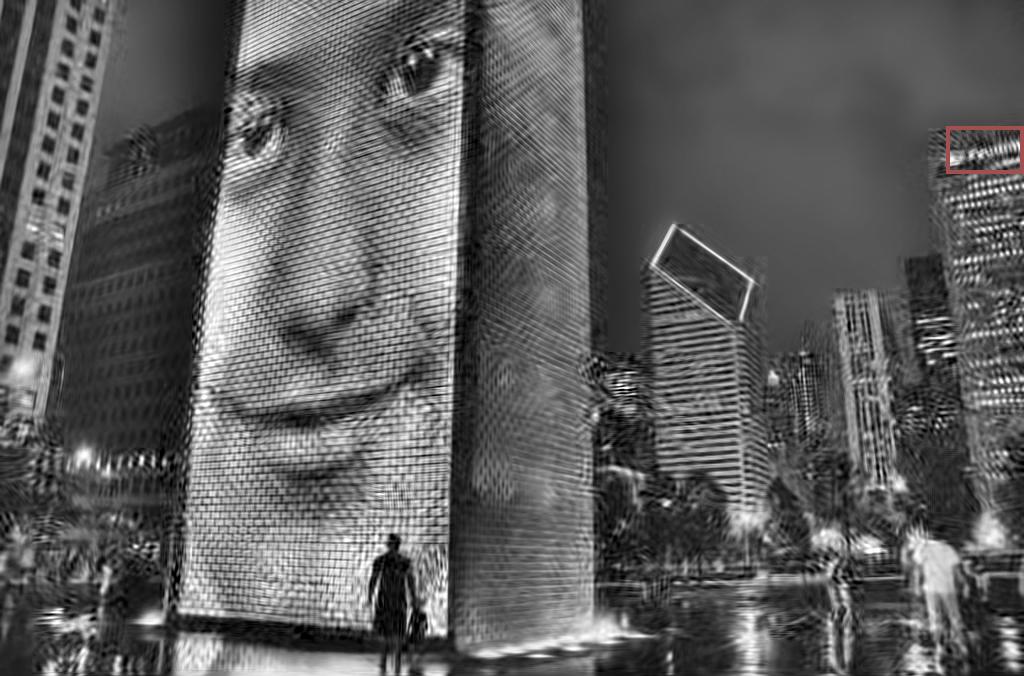}}
            \vspace{2pt}
            \centerline{\includegraphics[width=1\linewidth]{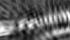}}
            \centerline{\scriptsize{23.38/0.7614}}
        \end{minipage}
        \begin{minipage}[b]{0.12\linewidth}
            \centerline{\scriptsize{CSNet$^+$~\cite{CSNet+_2020}}}
            \vspace{2pt}
            \centerline{\includegraphics[width=1\linewidth]{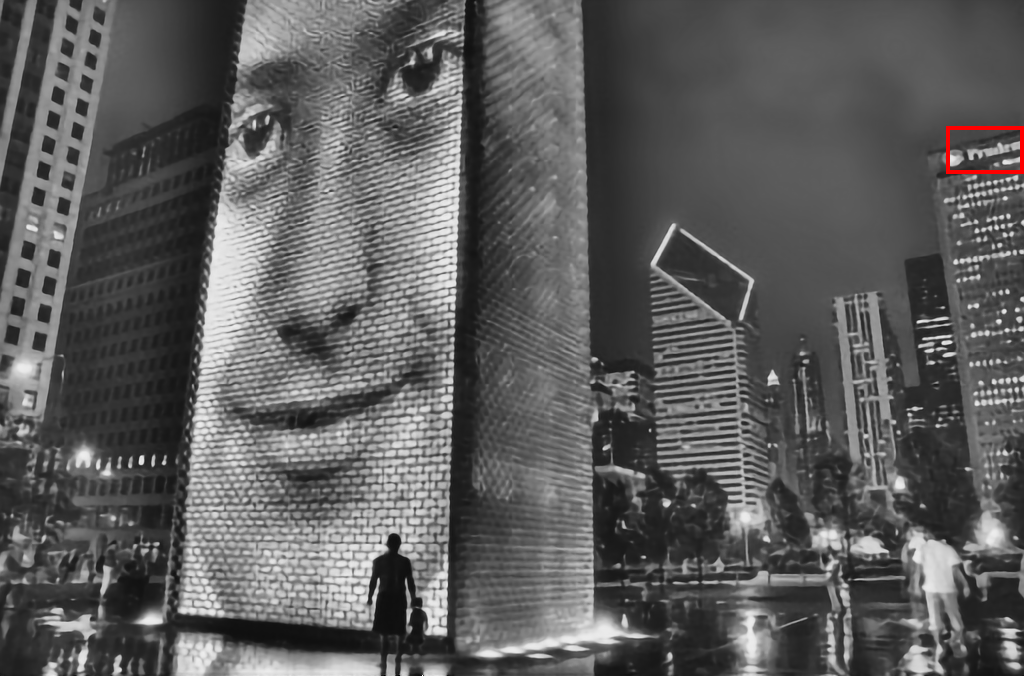}}
            \vspace{2pt}
            \centerline{\includegraphics[width=1\linewidth]{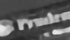}}
            \centerline{\scriptsize{24.55/0.8085}}
        \end{minipage}
        \begin{minipage}[b]{0.12\linewidth}
            \centerline{\scriptsize{AMP-Net~\cite{AMP-Net-2021}}}
            \vspace{2pt}
            \centerline{\includegraphics[width=1\linewidth]{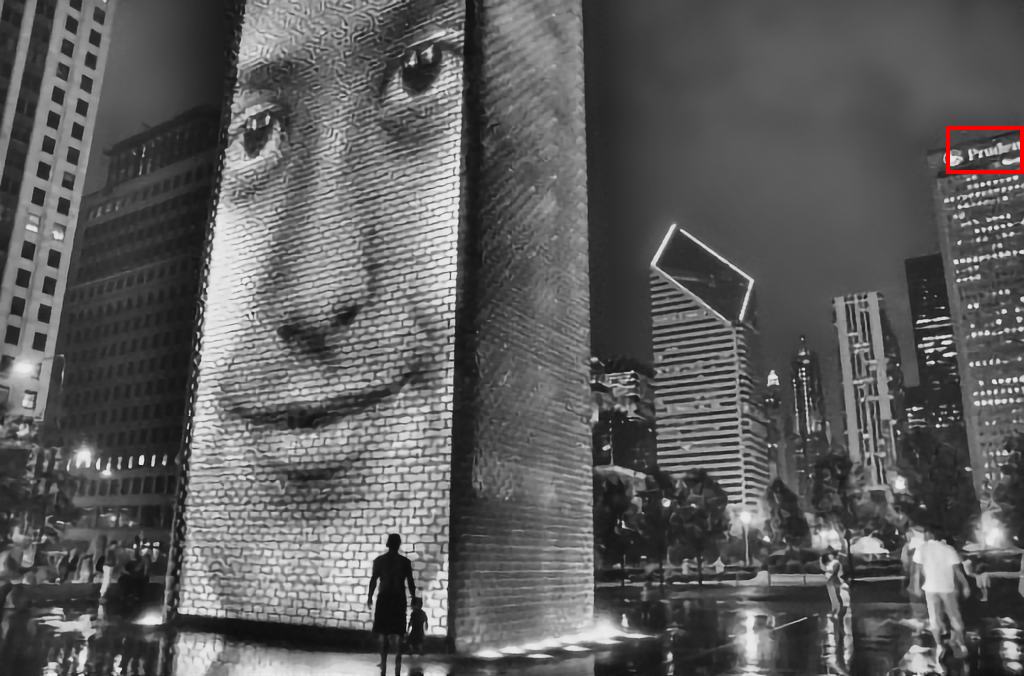}}
            \vspace{2pt}
            \centerline{\includegraphics[width=1\linewidth]{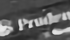}}
            \centerline{\scriptsize{24.82/0.8208}}
        \end{minipage}
        \begin{minipage}[b]{0.12\linewidth}
            \centerline{\scriptsize{ISTA-Net$^+$$^+$~\cite{you2021ISTA}}}
            \vspace{2pt}
            \centerline{\includegraphics[width=1\linewidth]{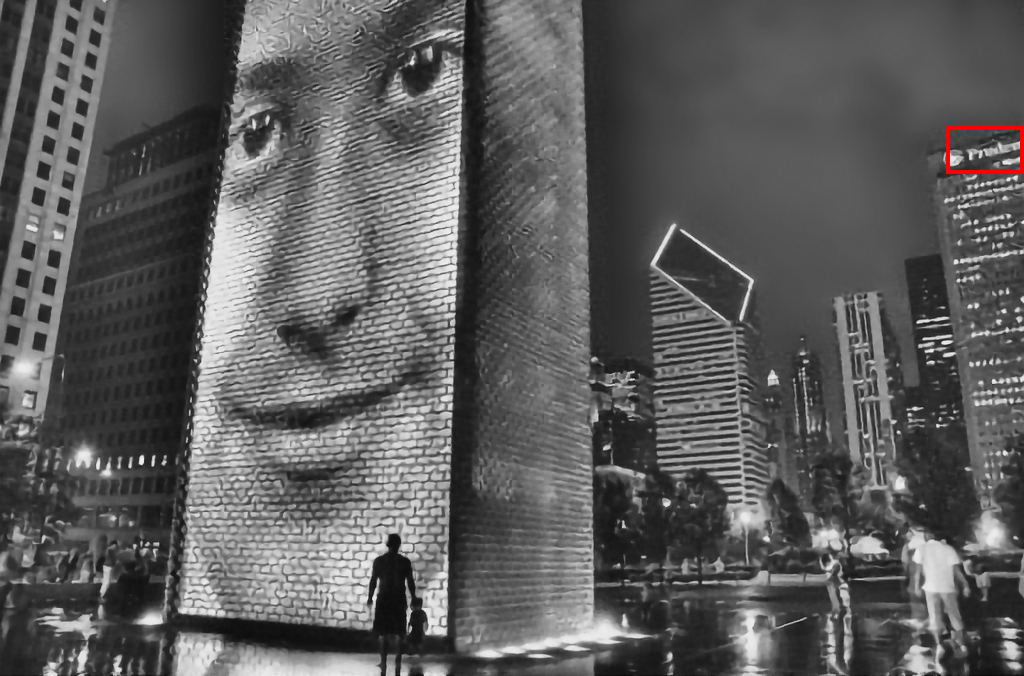}}
            \vspace{2pt}
            \centerline{\includegraphics[width=1\linewidth]{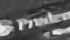}}
            \centerline{\scriptsize{24.48/0.8077}}
        \end{minipage}
        
        \vspace{10pt}

        \begin{minipage}[b]{0.12\linewidth}
            \centerline{\scriptsize{SCSNet$^+$~\cite{SCSNet-2019}}}
            \vspace{2pt}
            \centerline{\includegraphics[width=1\linewidth]{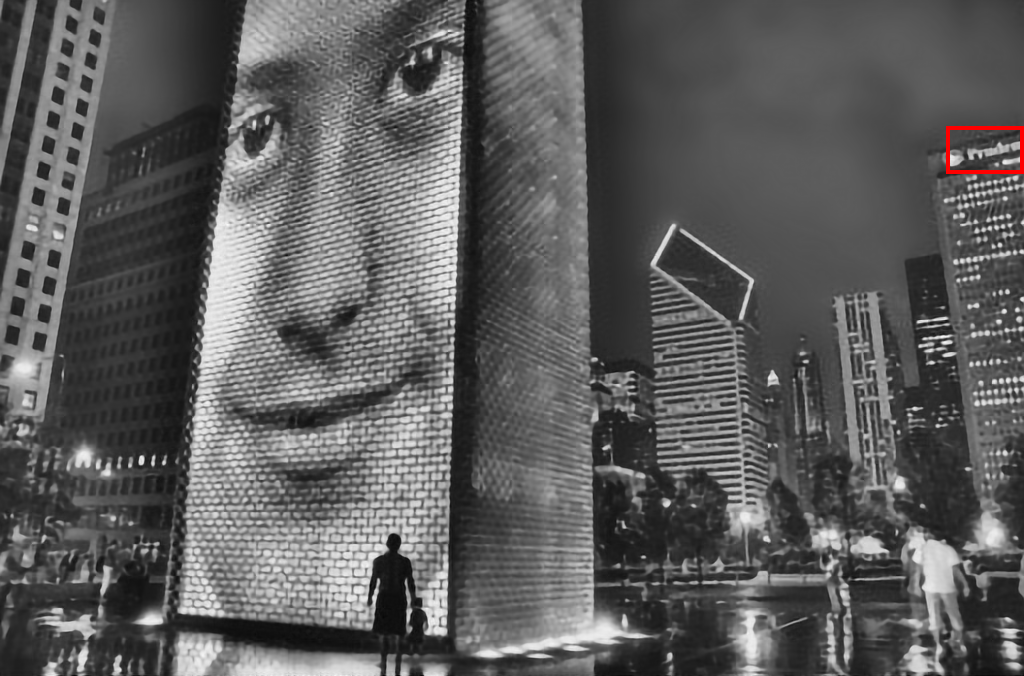}}
            \vspace{2pt}
            \centerline{\includegraphics[width=1\linewidth]{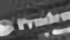}}
            \centerline{\scriptsize{24.60/0.8091}}
        \end{minipage}
        \begin{minipage}[b]{0.12\linewidth}
            \centerline{\scriptsize{OCTUF~\cite{song2023optimization_OCTUF}}}
            \vspace{2pt}
            \centerline{\includegraphics[width=1\linewidth]{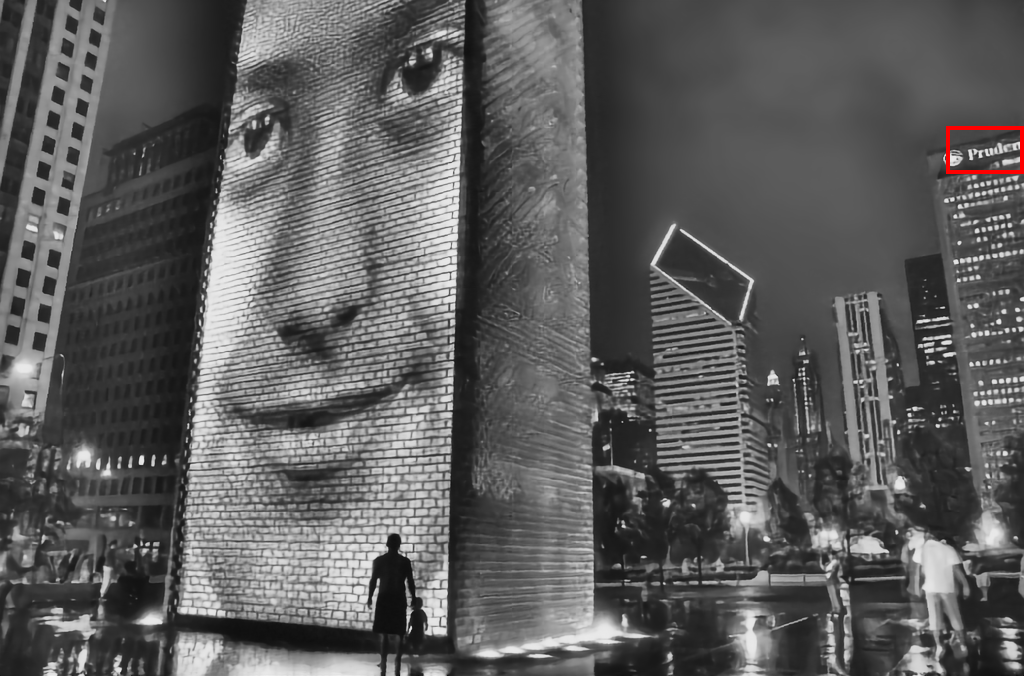}}
            \vspace{2pt}
            \centerline{\includegraphics[width=1\linewidth]{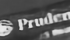}}
            \centerline{\scriptsize{25.80/0.8438}}
        \end{minipage}
        \begin{minipage}[b]{0.12\linewidth}
            \centerline{\scriptsize{TransCS~\cite{shenTransCSTransformerBasedHybrid2022_TransCS}}}
            \vspace{2pt}
            \centerline{\includegraphics[width=1\linewidth]{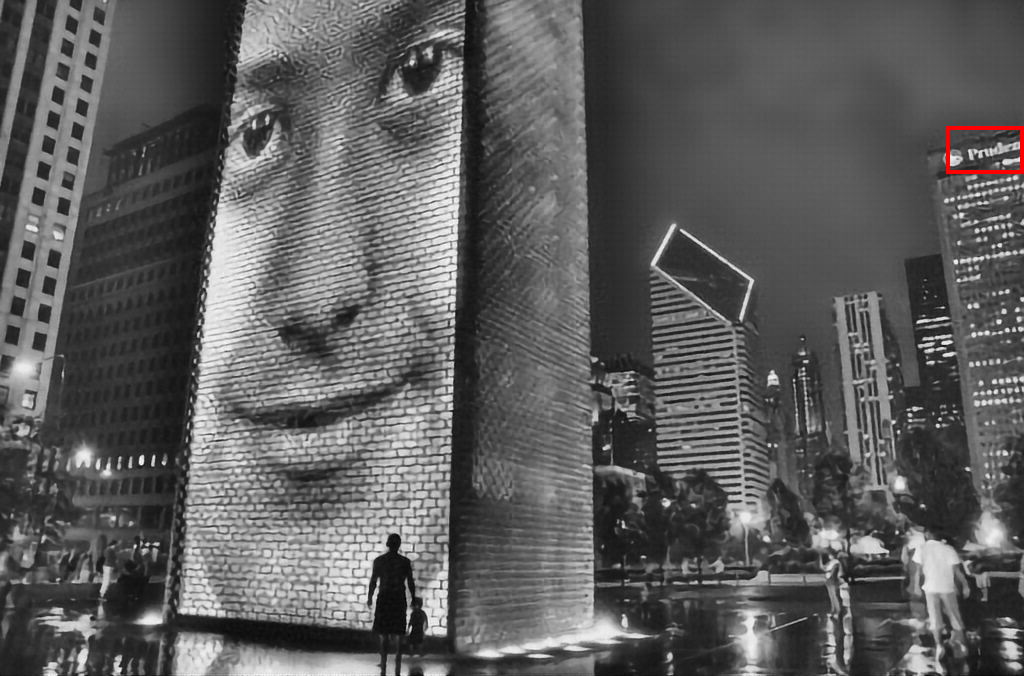}}
            \vspace{2pt}
            \centerline{\includegraphics[width=1\linewidth]{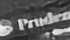}}
            \centerline{\scriptsize{24.90/0.8157}}
        \end{minipage}
        \begin{minipage}[b]{0.12\linewidth}
            \centerline{\scriptsize{\response{CSformer}~\cite{CSformer-2023}}}
            \vspace{2pt}
            \centerline{\includegraphics[width=1\linewidth]{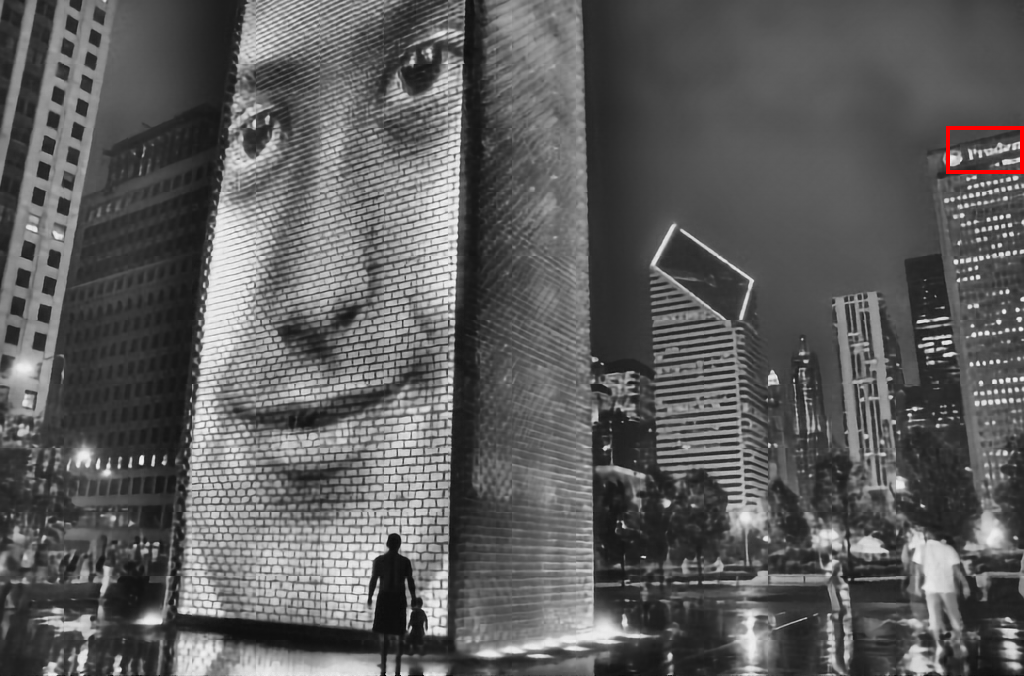}}
            \vspace{2pt}
            \centerline{\includegraphics[width=1\linewidth]{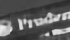}}
            \centerline{\scriptsize{\response{26.10/0.8589}}}
        \end{minipage}
        \begin{minipage}[b]{0.12\linewidth}
            \centerline{\scriptsize{\response{DPC-DUN}~\cite{Song2023-DPC-DUN}}}
            \vspace{2pt}
            \centerline{\includegraphics[width=1\linewidth]{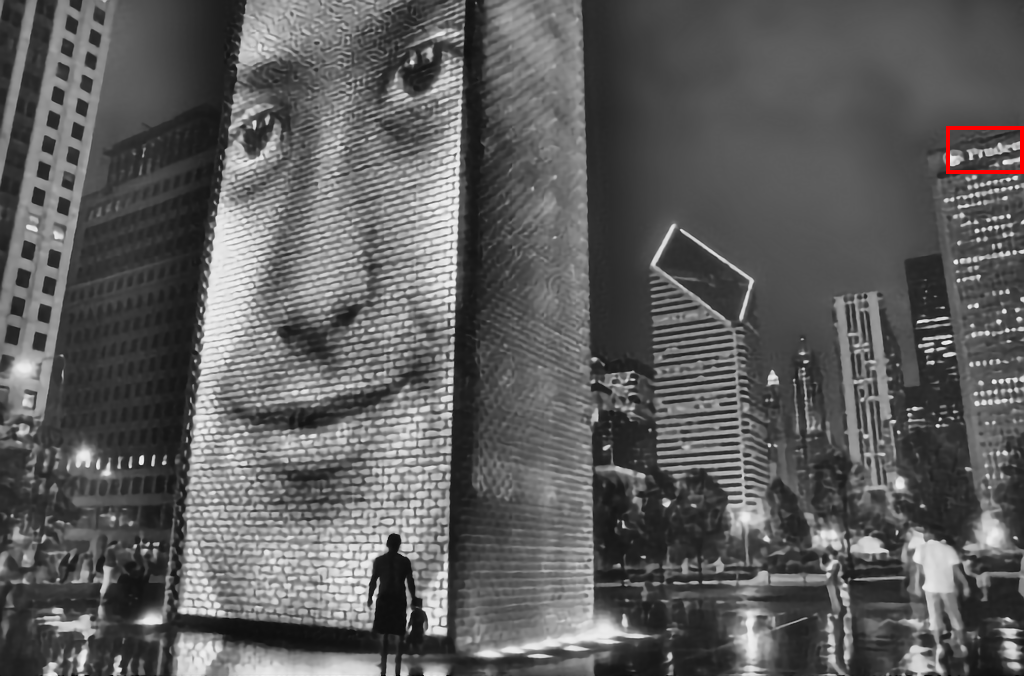}}
            \vspace{2pt}
            \centerline{\includegraphics[width=1\linewidth]{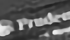}}
            \centerline{\scriptsize{\response{24.80/0.8234}}}
        \end{minipage}
        \begin{minipage}[b]{0.12\linewidth}
        \centerline{\scriptsize{\response{Ours-Non-Adaptive}}}
        \vspace{2pt}
        \centerline{\includegraphics[width=1\linewidth]{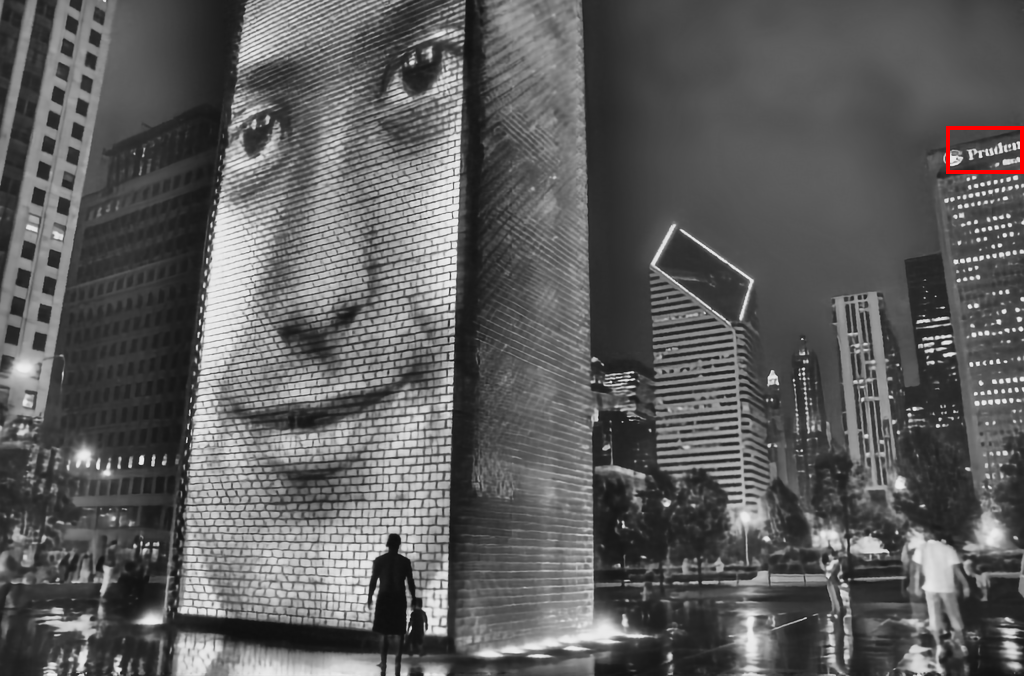}}
        \vspace{2pt}
        \centerline{\includegraphics[width=1\linewidth]{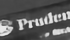}}
        \centerline{\scriptsize{\response{26.92/0.8760}}}
    \end{minipage}
        \begin{minipage}[b]{0.12\linewidth}
            \centerline{\scriptsize{\textbf{Uformer-ICS}$^+$}}
            \vspace{2pt}
            \centerline{\includegraphics[width=1\linewidth]{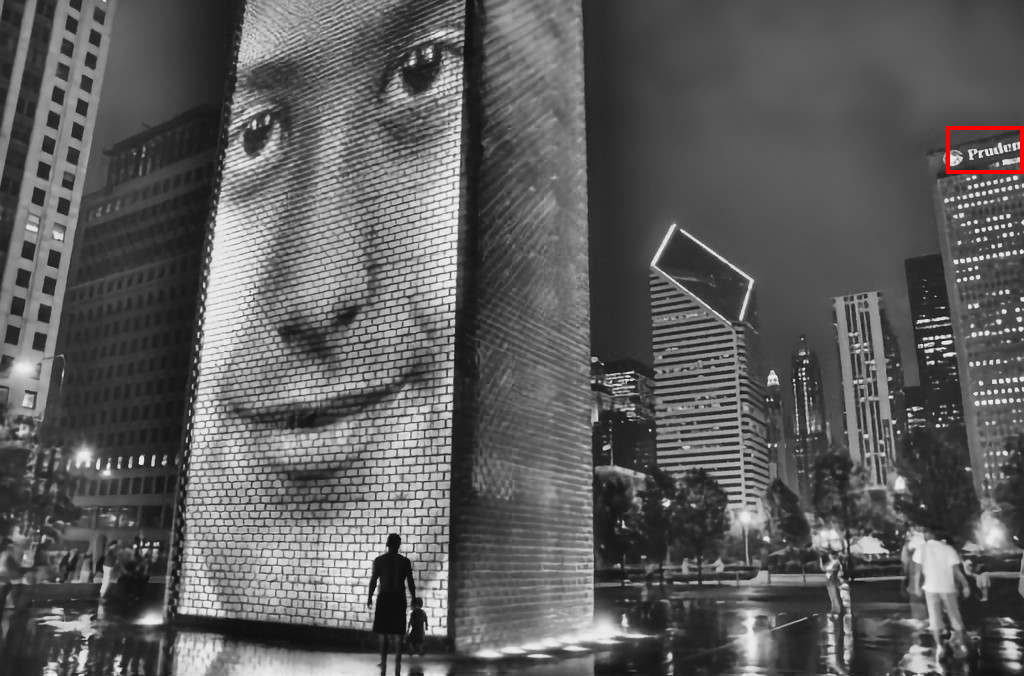}}
            \vspace{2pt}
            \centerline{\includegraphics[width=1\linewidth]{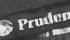}}
            \centerline{\scriptsize{\textcolor{blue}{26.99/0.8670}}}
        \end{minipage}
        \begin{minipage}[b]{0.12\linewidth}
            \centerline{\scriptsize{\textbf{Uformer-ICS}}}
            \vspace{2pt}
            \centerline{\includegraphics[width=1\linewidth]{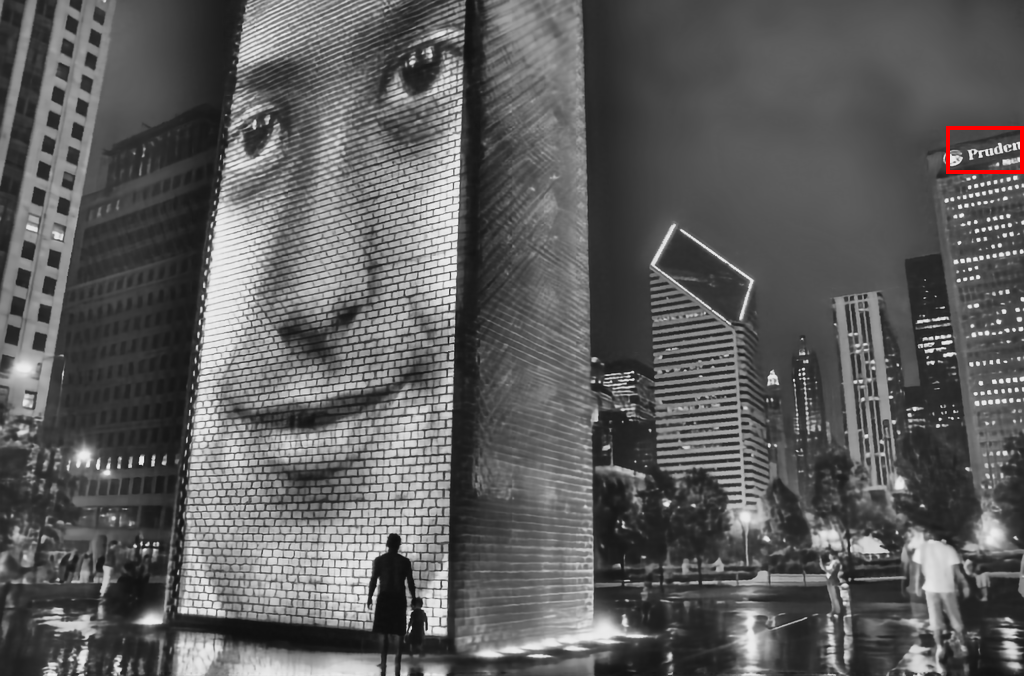}}
            \vspace{2pt}
            \centerline{\includegraphics[width=1\linewidth]{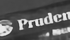}}
            \centerline{\scriptsize{\textbf{\textcolor{blue}{27.56/0.8804}}}}
        \end{minipage}
    \end{minipage}
    
    \centerline{\scriptsize{(b) Reconstruction results of the image ``img-076'' at $sr$ = 0.1}}    
      \vspace{-6pt}
    \caption{Reconstruction results of various CS methods at sampling ratios ($sr$s) of 0.04 and 0.1. Note that ``Ours-Non-Adaptive'' refers to the version of our Uformer-ICS without adaptive sampling module.}
    \label{Fig.visual_comparison}
\end{figure*}

\begin{table*}[!ht]
\centering
\caption{Time complexity comparison between different CS methods. The average PSNR scores and running times are listed for different image sizes, and the FLOPs is measured on an image of size $256\times 256$.}
  \vspace{-6pt}
\renewcommand{\arraystretch}{1.1}
\label{Tab.model_complexity}
\begin{tabular}{clllllcccccccc}
\toprule
\multirow{2}{*}{} & \multirow{2}{*}{Methods} & \multicolumn{4}{c}{PSNR (dB) / Times (s)} & \multirow{2}{*}{\tabincell{c}{Scalable\\Sampling}} & \multirow{2}{*}{Param.} & \multirow{2}{*}{FLOPs} \\ \cmidrule(lr){3-6}
 &  & $128\times 128$ & $256\times 256$ & $512\times 512$ & $1024\times 1024$ &  &  \\ \midrule
\multirow{4}{*}{\tabincell{c}{Traditional \\ CS methods}} 
& TV~\cite{TV-2009}                 & 22.78/0.2059 & 24.62/0.6852 & 27.91/2.6385 & 32.44/9.9298 & \Checkmark & -  & -  \\
& BM3D-DAMP~\cite{DAMP-2016}        & 21.67/3.3867 & 23.48/11.751 & 26.98/43.948 & 34.54/174.35 & \Checkmark & -  & -  \\
& MH-BCS-SPL~\cite{MH-2011}         & 22.89/0.6402 & 25.14/2.9934 & 29.70/20.629 & 38.26/108.15 & \Checkmark & -  & -  \\
& BCS-FOCUSS~\cite{BCS-FOCUSS_2017} & 22.63/1.2723 & 24.61/2.3107 & 28.78/7.4328 & 36.71/29.856  & \Checkmark & -  & -  \\
 \midrule
\multirow{9}{*}{\tabincell{c}{Deep \\ learning-based \\ CS methods}}
& CSNet$^{+}$~\cite{CSNet+_2020}      & 24.91/\textbf{0.0011} & 27.17/\textbf{0.0014} & 32.33/\textbf{0.0012} & 40.92/\textbf{0.0016} & $\times$ &0.89M & 24.26 \\
& SCSNet~\cite{SCSNet-2019} &24.26/0.0015 & 26.31/0.0012 & 31.06/0.0016 & 42.33/0.0018 & \Checkmark & 1.98M & 12.26 \\
& ISTA-Net$^{++}$~\cite{you2021ISTA} & 24.97/0.0091 & 27.37/0.0091 & 32.64/0.0091 & 40.35/0.0114 & \Checkmark & 1.70M & 50.15 \\
& AMP-Net~\cite{AMP-Net-2021}         & 25.32/0.0063 & 27.83/0.0063 & 33.33/0.0071 & 41.34/0.0094 & $\times$ & 0.86M & 25.77 \\
& OCTUF~\cite{song2023optimization_OCTUF} & 25.35/0.0212 & 28.15/0.0218 & 34.11/0.0235 & 43.68/0.0600 & $\times$ & 0.55M & 21.73 \\
& TransCS~\cite{shenTransCSTransformerBasedHybrid2022_TransCS} & 24.81/0.0216 & 27.04/0.0214 & 32.27/0.0237 & 39.12/0.0274 & $\times$ & 1.75M & 32.72 \\
& \response{DPC-DUN}~\cite{Song2023-DPC-DUN} & \response{25.33/0.0097} & \response{27.85/0.0092} & \response{33.45/0.0096} & \response{42.05/0.0117} &\response{$\times$} & \response{0.92M} & \response{42.67} \\
& \response{CSformer}~\cite{CSformer-2023} & \response{25.17/0.0128} & \response{27.65/0.0121} & \response{33.27/0.0131} & \response{42.88/0.0158} & \response{$\times$} & \response{6.67M} & \response{20.82}\\
& \textbf{Uformer-ICS$^+$} & \response{25.62/0.0252} & \response{28.56/0.0243} & \response{34.85/0.0292} & \response{44.63/0.0316} & \Checkmark & 9.82M & 49.83 \\
& \textbf{Uformer-ICS}  & \response{\textbf{25.62}/0.0253} & \response{\textbf{28.75}/0.0242} & \response{\textbf{35.23}/0.0273} & \response{\textbf{45.46}/0.0320} & $\times$ & 9.15M & 45.60 \\
 \bottomrule
\end{tabular}
\end{table*}

\subsubsection{Adaptive Sampling and Sparsity Estimation Methods}


A natural image usually contains different sparsities in different image blocks. To retain more information of the original image in the compressed measurements, we use an adaptive sampling strategy to sample image blocks based on their sparsity.
To estimate the block sparsity of the sampled image, we utilize three methods: SM, STD, and DIFF. Besides, to demonstrate the effectiveness of the adaptive sampling strategy, we build a non-adaptive model by discarding the adaptive sampling strategy from the proposed Uformer-ICS and train it by sampling each image block equally. 

Table~\ref{Tab.ablation_study_adaptive} shows the testing results of the adaptive and non-adaptive models on Set11 and Urban100 datasets at five sampling ratios $\{0.01, 0.04, 0.1, 0.25, 0.5\}$. As can be seen, our models using any sparsity estimation methods can obtain better reconstruction quality than the non-adaptive model. Compared to the non-adaptive model, the average PSNR results on Urban100 on five sampling ratios $\{0.01, 0.04, 0.1, 0.25, 0.5\}$ can be improved by 0.29 dB, 0.58 dB, and 0.80 dB for the adaptive models when using STD, DIFF, and SM, respectively.

In conclusion, the adaptive sampling strategy can effectively improve the reconstruction quality. 
Besides, the sparsity estimation method SM can lead to better reconstruction quality than STD and DIFF on the whole,  especially in the large dataset Urban100. This is because the DCT used by the SM method can allow it to accurately capture the image's energy distribution and exhibit greater robustness to noise compared to the STD and DIFF methods. Therefore, the SM method can better estimate the sparsity in our model. 
In the next sections, we use the SM to estimate sparsity by default if there are no special instructions.

\subsubsection{Multi-Channel Projection}

For the proposed reconstruction model, we construct the Uformer architecture to utilize the long-range dependency capture ability. Considering that CS characteristics are crucial for image reconstruction in CS tasks~\cite{OPINE-Net,AMP-Net-2021}, we develop a projection-based Transformer block to introduce the prior projection knowledge of CS into the transformer architecture. As shown in Fig.~\ref{Fig.mc-projection}, we integrate the MCP module into the original stacked Transformer blocks to construct the projection-based transformer block. Because the MCP reuses the measurement matrix and the measurements to update the image block, it can introduce the CS characteristics into the transformer architecture. Additionally, it brings in little computation overhead and few extra parameters by introducing only a learnable updating step vector. As can be seen from the settings (a) and (d) in Table~\ref{Tab.ablation_study}, the MCP only increases the model parameters from 9.1588 M to 9.1589 M. 

To evaluate the effectiveness of the MCP, we remove all the MCP modules in the projection-based transformer block to test the reconstruction performance. As shown in Table~\ref{Tab.ablation_study}, without the MCP modules, the PSNR scores will drop by approximately 0.1$\sim$1.17 dB and 0.24$\sim$2.47 dB on Set11 and Urban100 datasets, respectively, thereby demonstrating the importance of our MCP module for image reconstruction.

In addition,  Fig.~\ref{Fig.projection_res} presents the visualization results of the feature maps in the MCP modules of the proposed Uformer-ICS. The sampling ratio is set to 0.1. Note that we use the tail module $H_{tail}$ to convert the multi-channel feature maps into single-channel images for visualization. It can be observed that the projection results have finer details and clearer edges than the input feature maps.
The quantitative results in Table~\ref{Tab.ablation_study} and the visual results in Fig.~\ref{Fig.projection_res} verify the effectiveness of the multi-channel projection.

\subsubsection{Feature Fusion}
In the proposed model, we use the concatenation operation for feature fusion to aggregate the encoder features on the decoder side. In addition to the concatenation operation, another commonly used feature fusion method is the skip connection~\cite{he2016deep}, which directly adds the features of the two input branches element-wise. As can be seen from Table~\ref{Tab.ablation_study}, the reconstruction performances of the concatenation operation and skip connection have little difference on Set11 and Urban100, but the concatenation operation leads to a little better reconstruction performance on the whole. 
This is because the concatenation operation is more beneficial to generalization ability since it allows for more information preservation. In the following sections, we use the concatenation operation for feature fusion by default.


\subsubsection{Loss Function}

Our loss function contains three parts $\mathcal{L}_1$, $\mathcal{L}_2$ and $\mathcal{L}_3$. Besides two necessary loss items $\mathcal{L}_1$ and $\mathcal{L}_3$, $\mathcal{L}_2$ is a regularization term to ensure cycle consistency. To test the effectiveness of $\mathcal{L}_2$, we discard it in the training process and evaluate the performance. 
As can be seen from the settings (c) and (d) in Table~\ref{Tab.ablation_study}, the loss item $\mathcal{L}_2$ does benefit the reconstruction performance in most cases. Specifically, when the sampling ratio is larger than 0.1, the PSNR scores with $\mathcal{L}_2$ can improve by approximately 0.06$\sim$0.71 dB and 0.21$\sim$0.24 dB on Set11 and Urban100 datasets, respectively.

\subsection{Performance Evaluation}
We compare the proposed method with other CS methods in terms of the quantitative results, visual results, and model complexity. The PSNR and structural similarity index measure (SSIM) scores are used to quantitatively evaluate the reconstruction quality.

\subsubsection{Quantitative Comparison}
We calculate the average PSNR and SSIM scores on the five testing datasets at the sampling ratios ranging from 0.01--0.5. Table~\ref{Tab.Comparsion_CS} lists the experimental results for all competing CS methods. Benefiting from the powerful learning ability, all deep learning-based CS methods outperform the traditional CS methods. The proposed method achieves the best PSNR and SSIM scores for nearly all the sampling ratios. Specifically, the proposed Uformer-ICS can obtain significantly higher PSNR scores than other deep learning-based methods on the Urban100 dataset, which contains high-resolution images with different characteristics. The proposed Uformer-ICS can also achieve the best SSIM scores when the sampling ratio is smaller than 0.5. 

Additionally, we also calculate the average PSNR scores over all the five testing datasets at all sampling ratios, and the results of these deep learning-based CS methods are shown in Fig.~\ref{Fig.avg_PSNR}. It is evident that the proposed method achieve significantly higher scores than other deep learning-based methods for all sampling ratios. The quantitative comparison results presented in Table~\ref{Tab.Comparsion_CS} and Fig.~\ref{Fig.avg_PSNR} indicate that the proposed method outperforms all other CS methods and has a better generalization ability for different datasets.

Regarding scalable sampling and reconstruction ability, the proposed Uformer-ICS$^+$ requires only one-time training but can sample and reconstruct images at arbitrary sampling ratios.
Compared with the scalable CS methods SCSNet and ISTA-Net$^{++}$, the proposed Uformer-ICS$^+$ can still obtain better PSNR and SSIM scores in most cases. 
For the test datasets Set14 and BSD100, the proposed Uformer-ICS$^+$ achieve the best PSNR scores compared to all other non-scalable deep learning-based CS methods. These comparison results demonstrate the tremendous potential of the proposed method for scalable sampling and reconstruction.

\subsubsection{Visual Comparison} 
Fig.~\ref{Fig.visual_comparison} shows the visual results of different CS methods at the sampling ratios of 0.04 and 0.1, where ``Ours-Non-Adaptive'' refers to the version of our Uformer-ICS without the adaptive sampling module. The reconstruction results of all the traditional CS methods clearly show blocking artifacts, whereas those of the deep learning-based CS methods show significantly fewer blocking artifacts. Moreover, the visual results demonstrate that the proposed Uformer-ICS can recover more image details and clearer edges than other deep learning-based CS methods. For example, the reconstructed images of our Uformer-ICS exhibit more details around the letter area in the ``ppt3" and ``img-076" images. This is because the proposed sampling model uses an adaptive sampling strategy to allocate more sampling resources to the image block that has less sparsity, and the proposed reconstruction model can simultaneously capture the long-range dependency and local features of the image to reconstruct image information.

\subsection{Model Complexity}

We test the average running times on different image resolutions to evaluate the actual reconstruction efficiencies of all competing CS methods. Specifically, we randomly select five images from the ImageNet validation dataset and resize them to $128\times 128, 256\times 256, 512\times 512$, and $1024\times 1024$ using the bicubic interpolation algorithm. Thereafter, the average running times are calculated for the five images using these four kinds of image resolutions. Additionally, we utilize the number of floating-point operations (FLOPs) and parameters to theoretically quantify the time complexity of deep learning-based CS methods. It should be noted that we calculate the FLOPs on an input image of size $256\times 256$, and evaluate all the models at the sampling ratio of 0.25. The traditional CS methods are tested on the CPU device, whereas the deep learning-based CS methods are tested on the GPU device. 

Table~\ref{Tab.model_complexity} presents the comparison results of model complexity. 
As can be seen, traditional CS methods have much lower reconstruction quality and significantly lower running speeds, which may not satisfy the high-efficiency requirement of real-time applications. For example, they all cost more than two seconds to reconstruct the image of size $512\times 512$.
All deep learning-based CS methods have better PSNR scores and extremely faster speeds than traditional ones. Moreover, the running speeds of all deep learning-based CS methods are of the same order of magnitude. This is because the GPU device has a powerful computational ability and can run these deep learning-based models at high speed. The proposed method can achieve the best PSNR scores for all the image resolutions. 
Additionally, the Uformer-ICS$^+$ uses a single model for arbitrary sampling and reconstruction tasks, which can significantly reduce the number of parameters when sampling at multiple sampling ratios is required. Thus, the proposed method can well balance the trade-off between model complexity and reconstruction performance. It achieves the best reconstruction performance while maintaining a modest model complexity.

\section{Conclusion}
\label{Sec.conclusion}
 In this study, we proposed a novel transformer-based network for image CS, called Uformer-ICS, which effectively introduces two CS characteristics into the transformer architecture. We first designed an adaptive sampling architecture to retain the maximum possible information of the original image under a fixed sampling ratio. Specifically, it estimates sparsity from image measurements and linearly allocates measurement resources based on the estimated block sparsity. We utilized three sparsity estimation methods to evaluate the adaptive sampling architecture. Besides, we designed a MCP module to adapt the projection operation into the multi-channel feature domain. We constructed the projection-based transformer block by integrating the MCP module into the original transformer block and then built a symmetrical reconstruction model using the projection-based transformer blocks and residual convolutional blocks. Experimental results verified the effectiveness of the proposed Uformer-ICS, demonstrated its superiority over existing deep learning-based CS methods, and proved its tremendous potential for scalable sampling and reconstruction.

\ifCLASSOPTIONcaptionsoff
  \newpage
\fi

\section*{Acknowledge}
This work was supported in part by the National Natural Science Foundation of China under Grants 62071142 and 62001304, and by the Guangdong Basic and Applied Basic Research Foundation under Grant 2021A1515011406.

\normalem
\bibliographystyle{IEEEtran}
\bibliography{acmart}

\begin{IEEEbiography}[{\includegraphics[width=.95\linewidth]{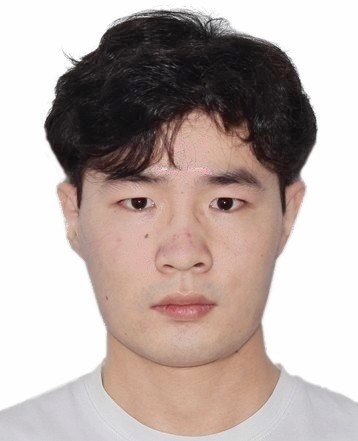}}]{Kuiyuan Zhang}
    received his M.S. degree in Computer Science and Technology from Harbin Institute of Technology, Shenzhen, China. He is currently pursuing the Ph.D degree in Computer Science and Technology from Harbin Institute of Technology, Shenzhen, China. His current research interests include image compressive sensing, image encryption and multimedia deepfake detection.  
\end{IEEEbiography}

\vskip 0pt plus -1fil

\begin{IEEEbiography}[{\includegraphics[width=.95\linewidth]{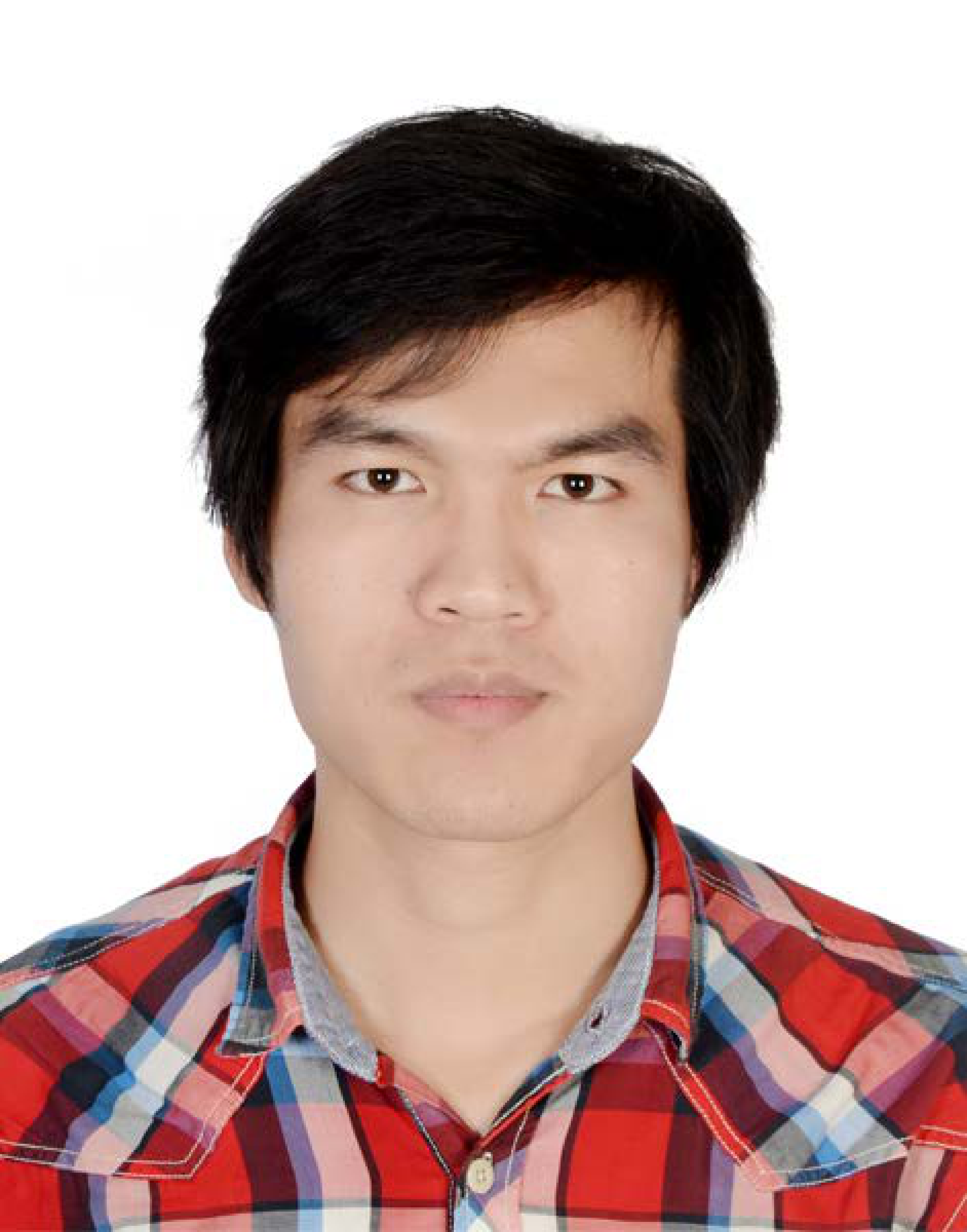}}]{Zhongyun Hua}
(Senior Member, IEEE) received the B.S. degree in software engineering from Chongqing University, Chongqing, China, in 2011, and the M.S. and Ph.D. degrees in software engineering from the University of Macau, Macau, China, in 2013 and 2016, respectively.

He is currently an Associate Professor with the School of Computer Science and Technology, Harbin Institute of Technology, Shenzhen, China. His works have appeared in prestigious venues such as \textit{IEEE Transactions on Dependable and Secure Computing, IEEE Transactions on Signal Processing, IEEE Transactions on Image Processing, IEEE Transactions on Cybernetics, IEEE Transactions on Systems, Man, and Cybernetics: Systems}, and \textit{ACM MultiMedia}.
His current research interests are focused on chaotic system, multimedia security, and secure cloud computing. He has published more than 70 papers on the subject, receiving more than 5000 citations. Dr. Hua has been recognized as a Highly Cited Researcher 2022. He is currently an Associate Editor for \textit{International Journal of Bifurcation and Chaos}. 
\end{IEEEbiography}

\vspace{-50pt}

\begin{IEEEbiography}
[{\includegraphics[width=1in,height=1.25in,clip,keepaspectratio]{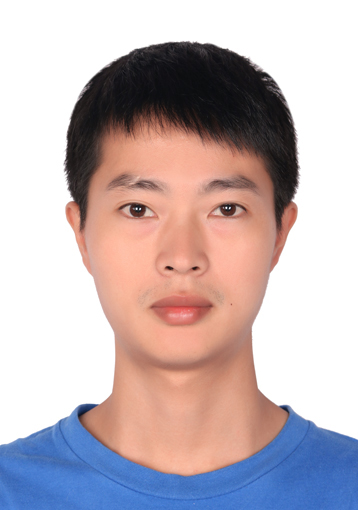}}]{Yuanman Li}
    (Member, IEEE) received the B.Eng. degree in software engineering from Chongqing University, Chongqing, China, in 2012, and the Ph.D. degree in computer science from the University of Macau, Macau, in 2018. 
    
    From 2018 to 2019, he was a Post-Doctoral Fellow with the State Key Laboratory of Internet of Things for Smart City, University of Macau. He is currently an Assistant Professor with the College of Electronics and Information Engineering, Shenzhen University, Shenzhen, China. His current research interests include data representation, multimedia security and forensics, computer vision, and machine learning.
\end{IEEEbiography}

\vspace{-50pt}

\begin{IEEEbiography}[{\includegraphics[width=.95\linewidth]{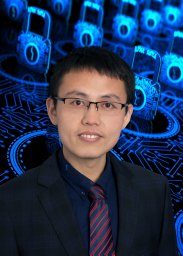}}]{Yushu Zhang}
 (Senior Member, IEEE) received the B.S. degree from the School of Science, North University of China, Taiyuan, China, in 2010, and the Ph.D. degree from the College of Computer Science and Technology, Chongqing University, Chongqing, China, in 2014. He held various research positions with the City University of Hong Kong, Southwest University, the University of Macau, and Deakin University. He is currently a Professor with the College of Computer Science and Technology, Nanjing University of Aeronautics and Astronautics, China. His research interests include multimedia security, blockchain, and artificial intelligence security. He is an Associate Editor of \textit{Information Sciences}, \textit{Journal of King Saud University-Computer and Information Sciences}, and \textit{Signal Processing}.

\end{IEEEbiography}

\vspace{-50pt}

\begin{IEEEbiography}
    [{\includegraphics[width=.95\linewidth]{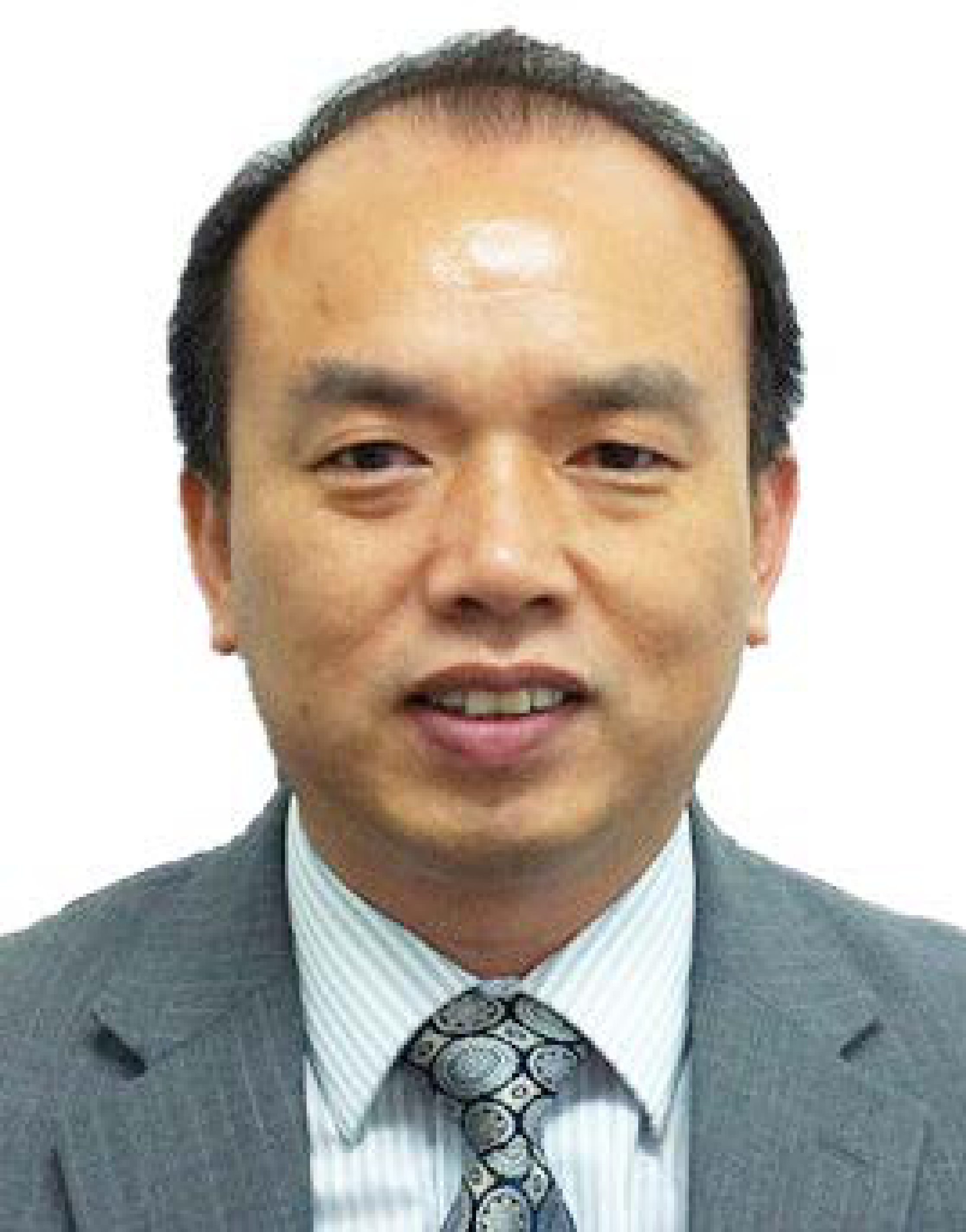}}]{Yicong Zhou}
(Senior Member, IEEE)  received the B.S. degree in electrical engineering from Hunan University, Changsha, China, in 1992, and the M.S. and Ph.D. degrees in electrical engineering from Tufts University, Medford, MA, USA, in 2008 and 2010, respectively.

He is a Professor with the Department of Computer and Information Science, University of Macau, Macau, China. His research interests include image processing, computer vision, machine learning, and multimedia security.

Dr. Zhou was recognized as one of the World's Top $2\%$ Scientists and one of the Highly Cited Researchers in 2020 and 2021. 
He received the Third Price of Macao Natural Science Award as a sole winner in 2020 and was a co-recipient in 2014. He has been a leading Co-Chair of the Technical Committee on Cognitive Computing of the IEEE Systems, Man, and Cybernetics Society since 2015. He serves as an Associate Editor for \textit{IEEE Transactions on Neural Networks and Learning Systems, IEEE Transactions
on Circuits and Systems for Video Technology}, and \textit{IEEE Transactions on GeoScience and Remote Sensing}.
\end{IEEEbiography}

\end{document}